\documentclass[12pt]{article}
\usepackage[utf8]{inputenc}
\usepackage{mathtools} % to use for equations 
\usepackage{apacite} %For APA referencing 
\usepackage{setspace}
\usepackage{amsmath}
\usepackage{xcolor} %add colors to texts
\usepackage{multirow}
\usepackage{float} %to put figure as they originally placed %way
\usepackage{graphicx} %to include figures
\usepackage{subcaption} %to include side-by-side figures
\usepackage{verbatim} % multi-line comments
\usepackage{algorithm} %To add algorithms
\usepackage{algpseudocode} %Also for algorithm adding
\usepackage{caption} %to center all captions, if use 
\usepackage{diffcoeff}  %partial div
\usepackage{setspace}
\usepackage[]{natbib} %et al format of citations
\usepackage{geometry}
\usepackage{chngcntr}
\counterwithin{figure}{section} % set numbering of equations and figures section Wise.
\counterwithin{equation}{section}
\usepackage{multicol} % add side by side equations
\usepackage{hyperref} %include hyperlinks to figures 
\usepackage{romannum}
\usepackage{titlesec}
\usepackage{tocloft} % to add headings without counting it
\usepackage{caption}
\usepackage{array}
\usepackage{booktabs}
\usepackage{orcidlink}
\usepackage{booktabs}
\usepackage{longtable}
\usepackage{rotating} %vertical tables

\usepackage{booktabs} %tbl wrapping in table
\usepackage{makecell} %tbl wrapping in table

\usepackage{chngcntr} %section wise tables
\counterwithin{table}{section} %section wise tables

\usepackage{authblk} % adding authors and affilations
\usepackage{array,multirow,graphicx}

\usepackage{subcaption}
\usepackage{tikz} % To sketch different graphics, such as grids
\usepackage{makecell}

\AtBeginDocument{\pagenumbering{arabic}} 
\title{A Marginal Maximum Likelihood Approach for Hierarchical Simultaneous Autoregressive Models with Missing Data}

\author{Anjana Wijayawardhana\thanks{Corresponding author: anjanaw@uow.edu.au} \orcidlink{0000-0003-3847-2671}}
\author{Thomas Suesse}
\author{David Gunawan}

\affil{School of Mathematics and Applied Statistics, University of Wollongong, Wollongong, Australia}

\date{}
\begin{document}
%\onehalfspacing

\maketitle
\renewcommand{\thefootnote}{\arabic{footnote}}
\begin{abstract}

%In the current literature, SAR estimation with missing data does not apply to extended SAR models, such as models with measurement errors. 
%However, under the presence of missing data, SAR estimation methods are less frequently discussed. It does not handle models with measurement errors.
    
Efficient estimation methods for simultaneous autoregressive (SAR) models with missing data in the response variable have been well-explored in the literature. A common practice is to introduce measurement error into SAR models to separate the noise component from the spatial process. However, prior research has not considered incorporating measurement error into SAR models with missing data. Maximum likelihood estimation for such models, especially with large datasets, poses significant computational challenges. This paper proposes an efficient likelihood-based estimation method, the marginal maximum likelihood (ML), for estimating SAR models on large datasets with measurement errors and a high percentage of missing data in the response variable. The spatial error model (SEM) and the spatial autoregressive model (SAM), two popular SAR model types, are considered. The missing data mechanism is assumed to follow a missing at random (MAR) pattern. We propose a fast method for marginal ML estimation with a computational complexity of $O(n^{3/2})$, where $n$ is the total number of observations. This complexity applies when the spatial weight matrix is constructed based on a local neighbourhood structure. The effectiveness of the proposed methods is demonstrated through simulations and real-world data applications.  

%Several computational approaches are developed to reduce the computational complexity of the proposed estimation methods. 

%in the marginal likelihood method and the EM method
%for any configuration of missing units,
%In this paper, we propose two efficient likelihood estimation methods: direct marginal maximum likelihood (ML) and expectation maximization (EM) algorithm for parameter estimation of two types of SAR models, spatial error model (SEM) and spatial autoregressive model (SAM) with measurement errors and missing data. We consider missing data and measurement errors that appear in the response variable. In addition, we consider the missing at random (MAR) missing data mechanism. To reduce the computational complexity of the proposed methods, several alternative computational approaches are discussed. An extensive theoretical and practical computational complexity study of proposed methods is discussed. To assess the proposed methods, simulated and real data sets are used. 

          
\end{abstract}
%\pagebreak

\section{Introduction}

Spatial regression models, or simultaneous autoregressive (SAR) models, extend traditional linear regression by accounting for spatial dependencies using a spatial weight matrix, $\textbf{W}$ (see Section~\ref{sec:models}). They are widely applied in fields such as ecology~\citep{tognelli2004analysis, ver2018spatial}, social sciences~\citep{angrist2004does, ammermueller2009peer}, criminology~\citep{glaeser1996crime}, and finance~\citep{longstaff2010subprime}. The two widely used SAR-type models are the spatial error model (SEM), where spatial dependence is in the error terms, and the spatial autoregressive model (SAM), where spatial dependence is in the dependent variable. See Section~\ref{sec:models} for further details.

% using the eigenvalues of $\textbf{W}$

The literature provides well-established estimation methods for SAR models with no missing values in the response variables. \citet{ord1975estimation} introduced an efficient maximum likelihood (ML) method. When dealing with sparse $\textbf{W}$, efficient sparse Cholesky factorisation algorithms are employed~\citep{pace1997sparse, pace1997performing, pace2004chebyshev} in ML estimation. Other methods include the method of moments (MOM)~\citep{kelejian1999generalized, kelejian2001asymptotic, lee2007gmm}, Bayesian approaches~\citep{hepple1979bayesian, anselin1988spatial, lesage1997bayesian}, and instrumental variable (IV) methods~\citep{lee2003best}.

Missing values in the response variable are common, and SAR models can be inaccurately estimated if $\textbf{W}$ is based only on the locations of observed responses, leading to biased results~\citep{wang2013estimation, benedetti2020spatial}. Various estimation methods are proposed to address this issue. \citet{lesage2004models} proposed an iterative algorithm similar to the expectation-maximization (EM) algorithm~\citep{dempster1977maximum}, using approximations to avoid computationally demanding matrix inversions. \citet{suesse2017computational} refined this approach and proposed a valid EM algorithm by incorporating exact terms. Alternatively, \citet{suesse2018marginal} developed a method that directly maximises the marginal log-likelihood of observed data, which is computationally faster and avoids the EM algorithm's convergence issues. Additionally, \citet{kelejian2010spatial} introduced an IV estimator for the SAM with missing responses, and \citet{luo2021ipw} proposed an inverse probability weighting (IPW) based robust estimator for the same model.

%For example, \citet{burden2015sar} and \citet{suesse2018estimation} used ML estimation for SAR models with measurement errors. \citet{bivand2015spatial} and \citet{gomez2021estimating} employed the Integrated Nested Laplace Approximation (INLA)~\citep{rue2009approximate} for SAR models with measurement errors. These studies, however, assume no missing values in the response variable. 

In spatial statistics, the observations are noisy measurements of the underlying spatial unobserved latent process. As a result, a measurement error is usually added to SAR models. 
\citet{burden2015sar} and \citet{suesse2018estimation} used ML estimation for SAR models with measurement errors, while \citet{bivand2015spatial} and \citet{gomez2021estimating} employed the Integrated Nested Laplace Approximation (INLA)~\citep{rue2009approximate} for such models. These studies, however, assume no missing values in the response variable.

{Our article makes two contributions. First, we introduce hierarchical SAR (H-SAR) models that account for measurement errors and missing data in the response variable. Second, we introduce a novel marginal ML method along with two alternative computational approaches that significantly reduce its computational complexity. Our most efficient approach, called the parameterisation approach, reduces the complexity of the proposed marginal ML method from $O(n^3)$ to $O(n^{3/2})$, where $n$ is the total number of observations; see Section \ref{sec:compaccept}
for further details.  We illustrate the H-SAR models and their marginal ML method empirically, using simulated and real data.}

The remainder of this paper is organised as follows. Section~\ref{sec:models} discusses the hierarchical SAR models. Section~\ref{sec:estimation} presents the marginal ML method. Efficient computational strategies to reduce the complexity of the proposed algorithm are also discussed. 
In Section ~\ref{sec:simulation}, we evaluate the performance of the estimation method using simulated datasets. 
Section ~\ref{sec:example} discusses the real data application. Section ~\ref{sec:conclusion} concludes.
The paper has an online supplement containing some further technical and empirical results.
%\textcolor{red}{add links to the supplementary and the codes}

%by conducting simulations with diverse missing data configurations

%model is
%\begin{equation}
%\label{eqn:trueprocess}
% {y}_i= z_i+ {\epsilon}_i ~~~~~i=1,2, \ldots, n,
%\end{equation}   
%\noindent where ${z}_i$ represents the unobserved, spatial latent process, and $\epsilon_i$ is the additive measurement error at the spatial location $s_i$. We assume that $\epsilon_i$ follows a normal distribution with a mean of $0$ and a variance  $\sigma_{\epsilon}^2$.

%We consider the hierarchical spatial autoregressive model (H-SAM) by replacing the hidden spatial process $z_i$ (or $\textbf{z}$ in vector notation) in Equation~\eqref{eqn:trueprocess} by the standard SAM,
%\begin{equation}
%\label{eq:SAM}
% \begin{split}
%        \textbf{y}&=\textbf{z}+\boldsymbol{\epsilon},\\
%     \textbf{z}&=\textbf{X}\boldsymbol{\beta}+\rho \textbf{W}\textbf{z}+\textbf{e}, 
% \end{split}  
%\end{equation}

%. $\textbf{z}=(z_1,z_2, \hdots, z_n)^\top$ is the vector of latent response variables at $n$ spatial locations $s_1, \hdots,s_n$

\section{Hierarchical Simultaneous Autoregressive  models \label{sec:models}}

In this section, we present hierarchical simultaneous autoregressive (H-SAR) models. In many practical applications, observations are a noisy reflection of the underlying scientific process, so measurement errors are commonly included in the spatial statistical model.
Given a vector of observed spatial data $\textbf{y}=(y_1, y_2, \ldots, y_n)^\top$, and a vector of latent spatial processes $\textbf{z}=(z_1, z_2, \ldots, z_n)^\top$ at $n$ spatial locations $s_1, \ldots, s_n$, the 
hierarchical spatial autoregressive model (H-SAM) is given by
\begin{equation}
\label{eq:SAM}
 \begin{split}
        \textbf{y}&=\textbf{z}+\boldsymbol{\epsilon},\\
     \textbf{z}&=\textbf{X}\boldsymbol{\beta}+\rho \textbf{W}\textbf{z}+\textbf{e}, 
 \end{split}  
\end{equation}
and the hierarchical spatial error model (H-SEM) is given by
\begin{equation}
\label{eq:SEM}
\begin{split}
\textbf{y}&=\textbf{z}+\boldsymbol{\epsilon},\\
 \textbf{z}&=\textbf{X}\boldsymbol{\beta}+\textbf{u},\\
 \textbf{u}&=\rho \textbf{W}\textbf{u}+\textbf{e} ,
\end{split}
\end{equation}
\noindent where $\boldsymbol{\epsilon}=(\epsilon_{1}, \epsilon_{2}, \ldots, \epsilon_{n})^\top$ is a vector of additive measurement errors, assumed to follow a multivariate normal distribution with the mean vector $\boldsymbol{0}$ and covariance matrix  $\sigma_{\epsilon}^2\textbf{I}_n$, where $\sigma_{\epsilon}^2$ is a variance parameter and $\textbf{I}_n$ is the $n\times n$ identity matrix. The matrix $\textbf{X}$ is the $n\times (r+1)$ matrix of predictors/covariates, and $\textbf{W}$ is the $n\times n$ spatial weight matrix. The error term $\textbf{e}$ is assumed to follow a multivariate normal distribution with the mean vector $\textbf{0}$ and covariance 
matrix $\sigma^2_{\textbf{e}}\textbf{I}_n$, where $\sigma^2_{\textbf{e}}$ is a variance parameter. The vector of fixed effect parameters is denoted by $\boldsymbol{\beta}=(\beta_0, \hdots, \beta_r)^\top$, and $\rho$ is the spatial autocorrelation parameter~\citep{anselin1988spatial,allison2001missing,lesage2009introduction}. We assume that $\textbf{W}$ is sparse, as commonly observed in numerous real-world scenarios. However, this may not necessarily be the case.

%Let $W_{ij}$ be the $i^{th}$ row and $j^{th}$ column entry of the spatial weight matrix $\textbf{W}$. The entry $W_{ij}$ is non-zero if the unit $i$ is the neighbour of the unit $j$. By definition, the diagonal of the spatial weight matrix $\textbf{W}$ is zero, i.e. $W_{ii}=0$. Several strategies for constructing $\textbf{W}$  have been proposed in the literature (see ~\citet{ ord1975estimation,anselin1988spatial, kelley1997fast} for further details). The spatial weight matrix $\textbf{W}$ has the dimension of $n \times n$, where $n$ is the number of observations. 
%Section \ref{sub.sec:complexity} provides a detailed explanation of the methodology used to construct the spatial weight matrices that are employed in the simulation studies in this paper.

%Since the error vector $\textbf{e}$ follows a normal distribution, the latent process $\textbf{z}$ is normally distributed with mean vector $\boldsymbol{\mu}_{\textbf{z}}$ and covariance matrix $\boldsymbol{\Sigma}_{\textbf{z}}$. 

Assuming that the measurement error $\boldsymbol{\epsilon}$ and the error $\textbf{e}$ vectors are normally distributed, the response variable $\textbf{y}$ in both H-SEM and H-SAM is normally distributed with mean vector $\boldsymbol{\mu}$ and covariance matrix $\boldsymbol{\Sigma}$. Table~\ref{tbl:properties} provides the mean vectors and covariance matrices for the H-SAR models being considered.

\begin{table}[h]
\caption{Expressions for mean ($\boldsymbol{\mu}$), covariance ($\boldsymbol{\Sigma}$), and $\textbf{V}$ for the H-SAR models with $\textbf{A}=\textbf{I}_n-\rho\textbf{W}$, $\omega= \sigma^2_{\epsilon}$ and $\theta=\sigma^2_{\textbf{e}}/\sigma^2_{\epsilon}$.}
\label{tbl:properties}
\centering

\begin{tabular}{ccc}
Terms               & H-SAM & H-SEM \\ 
\cline{1-3}
$\boldsymbol{\mu}$                & $\textbf{A}^{-1}\textbf{X}\boldsymbol{\beta}$    &  $\textbf{X}\boldsymbol{\beta}$ \\
$\boldsymbol{\Sigma}$ &  $\omega(\textbf{I}_n+\theta (\textbf{A}^\top\textbf{A})^{-1})$    &  $\omega(\textbf{I}_n+\theta (\textbf{A}^\top\textbf{A})^{-1})$   \\
$\textbf{V}$  & $(\textbf{I}_n+\theta (\textbf{A}^\top\textbf{A})^{-1})$  &$(\textbf{I}_n+\theta (\textbf{A}^\top\textbf{A})^{-1})$ \\
%$\textbf{M}=\textbf{V}^{-1}$ & $(\textbf{I}_n+\theta (\textbf{A}^\top\textbf{A})^{-1})^{-1}$  & $(\textbf{I}_n+\theta (\textbf{A}^\top\textbf{A})^{-1})^{-1}$
\end{tabular} 
\end{table}

%\begin{table}[h]
%\caption{Expressions for mean ($\boldsymbol{\mu}$), covariance ($\boldsymbol{\Sigma}$), $\textbf{V}$, and $\textbf{M}$ for SAR and H-SAR models with $\textbf{A}=\textbf{I}_n-\rho\textbf{W}$, $\omega= \sigma^2_{\epsilon}$ and $\theta=\sigma^2_{\textbf{e}}/\sigma^2_{\epsilon}$.}
%\label{tbl:properties}
%\centering

%\begin{tabular}{c|cc|cc}
%Term               & SAM &   SEM & H-SAM & H-SEM \\ 
%\cline{1-5}
%$\boldsymbol{\mu}$              & $\textbf{A}^{-1}\textbf{X}\boldsymbol{\beta}$    &  $\textbf{X}\boldsymbol{\beta}$  & $\textbf{A}^{-1}\textbf{X}\boldsymbol{\beta}$    &  $\textbf{X}\boldsymbol{\beta}$ \\
%$\boldsymbol{\Sigma}$ & $\sigma^2_{\textbf{e}}(\textbf{A}^\top\textbf{A})^{-1}$    &  $\sigma^2_{\textbf{e}}(\textbf{A}^\top\textbf{A})^{-1}$ & $\omega(\textbf{I}_n+\theta (\textbf{A}^\top\textbf{A})^{-1})$    &  $\omega(\textbf{I}_n+\theta (\textbf{A}^\top\textbf{A})^{-1})$   \\
%$\textbf{V}$ & $(\textbf{A}^\top\textbf{A})^{-1}$ & $(\textbf{A}^\top\textbf{A})^{-1}$  & $(\textbf{I}_n+\theta (\textbf{A}^\top\textbf{A})^{-1})$  &$(\textbf{I}_n+\theta (\textbf{A}^\top\textbf{A})^{-1})$ \\
%$\textbf{M}=\textbf{V}^{-1}$ & $\textbf{A}^\top\textbf{A}$  & $\textbf{A}^\top\textbf{A}$& $(\textbf{I}_n+\theta (\textbf{A}^\top\textbf{A})^{-1})^{-1}$  & $(\textbf{I}_n+\theta (\textbf{A}^\top\textbf{A})^{-1})^{-1}$
%\end{tabular} 
%\end{table}

To ensure a valid covariance matrix in H-SAM and H-SEM, the  spatial autocorrelation parameter \(\rho\) must not take the values \ $\frac{1}{\lambda_{(1)}}, \frac{1}{\lambda_{(2)}}, \ldots, \frac{1}{\lambda_{(n)}}$, where $\lambda_{(1)}, \lambda_{(2)}, \ldots, \lambda_{(n)}$ are the eigenvalues of $\textbf{W}$  in ascending order~\citep{li2012one}. When $\textbf{W}$ is normalised either by row or column, $\rho$ is constrained to $\frac{1}{\lambda_{(1)}} < \rho < 1$~\citep{lesage2009introduction}. 
The log-likelihood function of $\textbf{y}$ in terms of the model parameters $\boldsymbol{\phi} = (\boldsymbol{\beta}^\top, \rho, \omega, \theta)^\top$ for H-SAR models is

\begin{equation}
\label{com_like_2}
   \textrm{log}f(\textbf{y};\omega, \theta,\rho,\boldsymbol{\beta})=-\frac{n}{2}\textrm{log}(2\pi)-\frac{n}{2}\textrm{log}(\omega)-\frac{1}{2}\textrm{log}|\textbf{V}|-\frac{1}{2\omega}\textbf{r}^\top\textbf{V}^{-1}\textbf{r},
\end{equation}
\noindent where $\textbf{r}=\textbf{y}-\boldsymbol{\mu}$ is the vector of residuals. Expressions for the $\textbf{V}$ and $\boldsymbol{\mu}$ are given in Table~\ref{tbl:properties}.
The next section discusses the proposed marginal ML estimation method for estimating the H-SAR models with missing responses under the missing at random (MAR) mechanism.

%~\citet{suesse2018estimation} developed an efficient ML estimation method for estimating H-SAR models with fully observed data. 

%The same constraints apply to $\rho $ in H-SAR models for a valid covariance matrix $\boldsymbol{\Sigma}$, as shown by \citet{suesse2018estimation}.

\section{H-SAR Models with Missing Data and their Marginal Maximum Likelihood Methods}
%\section{Hierarchical simultaneous autoregressive models under MAR and their Estimation}
\label{sec:estimation}
Section~\ref{sec:4.1} discusses H-SAR models with missing data in the response values. Section~\ref{sec:direct.marginal.likelihood} presents the marginal ML (MML) method, while Section~\ref{sec:compaccept} explores the computational aspects of the proposed MML method, presenting two computational approaches.

\subsection{Hierarchical Simultaneous Autoregressive models with randomly missing responses}
\label{sec:4.1}

%, where $n_o$ and $n_u$ are the numbers of observed and missing response variables, respectively

Let $\textbf{y}_o$ represent the subset of $\textbf{y}$ containing $n_o$ observed response variables, and $\textbf{y}_u$ represent the subset of $\textbf{y}$ containing $n_u$ missing response variables. The complete-data vector is denoted by $\textbf{y} = (\textbf{y}_o^\top, \textbf{y}_u^\top)^\top$. The matrices $\textbf{X}$ and $\textbf{W}$ are divided into distinct parts as follows:
\begin{equation}
\label{mat:portions_of_xw}
\textbf{X}=
\begin{pmatrix}
    \textbf{X}_o\\
   \textbf{X}_u
\end{pmatrix},~\textbf{W}=
\begin{pmatrix}
    \textbf{W}_{oo} &  \textbf{W}_{ou}\\
    \textbf{W}_{uo} & \textbf{W}_{uu}
\end{pmatrix},
\end{equation}

\noindent where $\textbf{X}_o$ and $\textbf{X}_u$ are the corresponding matrices of covariates for the observed and unobserved response variables, respectively, and $\textbf{W}_{oo}$, $\textbf{W}_{ou}$, $\textbf{W}_{uo}$, and $\textbf{W}_{uu}$ represent the sub-matrices of $\textbf{W}$. 

%corresponding to the relationships between the observed and unobserved units.
%Given missing data in the response variable, employing conventional methods such as deletion approaches is not suitable for SAR models.

%\begin{equation}
%\label{eq:dis_obs}
%    f(\textbf{y}_{o} ; \boldsymbol{\phi})=\int f(\textbf{y} ; \boldsymbol{\phi})d\textbf{y}_{u},
%\end{equation}
%\noindent where 

In this study, we assume that the missing responses, $\textbf{y}_u$, are missing at random (MAR). The MML method under the MAR mechanism involves maximising the marginal likelihood of the observed data, $\textbf{y}_o$; see~\cite{little2019statistical} for further details. To compute the marginal likelihood of $\textbf{y}_o$, the unobserved data $\textbf{y}_u$ must be integrated out from the complete data density of $\textbf{y}$, $f(\textbf{y}_{o} ; \boldsymbol{\phi})=\int f(\textbf{y} ; \boldsymbol{\phi})d\textbf{y}_{u}$, where
$f(\textbf{y} ;\boldsymbol{\phi})$ is the complete data density of $\textbf{y}$. 
In the following section, we derive the marginal distribution of $\textbf{y}_o$ using properties of the multivariate normal distribution and present the MML estimators.

\subsection{Marginal maximum likelihood estimation method} 
\label{sec:direct.marginal.likelihood}
\begin{comment}
    This means the marginal distribution of $\textbf{z}_o$ in Equation~\eqref{eq:dis_obs} can be written in closed form, and 
the sub-vector $\boldsymbol{\mu}_o$ and sub-matrix $\textbf{V}_{oo}$ are obtained by partitioning the mean vector $\boldsymbol{\mu}$ and the matrix $\textbf{V}$ as in ~\eqref{eq:dkjsd}.
\end{comment}

% In other words, $\textbf{z}_o \sim N(\boldsymbol{\mu}_o, \omega\textbf{V}_{oo})$

This section discusses the proposed MML estimation method. The vector $\boldsymbol{\mu}$ and the matrix $\textbf{V}$ in Table \ref{tbl:properties} are partitioned as 

\begin{equation}
    \label{eq:dkjsd}
    \boldsymbol{\mu}=
\begin{pmatrix}
    \boldsymbol{\mu}_o\\
   \boldsymbol{\mu}_u
\end{pmatrix},~\textbf{V}= \begin{pmatrix}
    \textbf{V}_{oo} &  \textbf{V}_{ou}\\
    \textbf{V}_{uo} & \textbf{V}_{uu}
\end{pmatrix}.
\end{equation}
Since $\textbf{y}$ is a multivariate normal random variable, $\textbf{y}_o$ also follows a multivariate normal distribution with the mean vector $\boldsymbol{\mu}_o$ and the covariance matrix $\boldsymbol{\Sigma}_{oo}=\omega\textbf{V}_{oo}$~\citep{petersen2008matrix}; see Equation~\eqref{eq:dkjsd}. To compute the log-likelihood of the marginal distribution of $\textbf{y}_o$, replace $\textbf{V}$ with $\textbf{V}_{oo}$, $\boldsymbol{\mu}$ with $\boldsymbol{\mu}_{o}$, $\textbf{y}$ with $\textbf{y}_{o}$, and $n$ with $n_o$ in Equation~\eqref{com_like_2}. This yields the following expression for the marginal log-likelihood of $\textbf{y}_o$:
\begin{equation} 
\label{eq:maginallik_me}
    \text{log}f(\textbf{y}_o;\omega,\theta,\rho,\boldsymbol{\beta})=-\frac{n_o}{2}\textrm{log}(2\pi)-\frac{n_o}{2}\textrm{log}(\omega)-\frac{1}{2}\textrm{log}|\textbf{V}_{oo}|-\frac{1}{2\omega}\textbf{r}_o^\top\textbf{V}^{-1}_{oo}\textbf{r}_o,
\end{equation} 
\noindent where $\textbf{r}_o=\textbf{y}_o-\boldsymbol{\mu}_o$.
Maximising Equation~\eqref{eq:maginallik_me} with respect to $\boldsymbol{\beta}$ and $\boldsymbol{\omega}$ (taking partial derivatives and setting to zero) while holding $\theta$ and $\rho$ fixed, we obtain the closed form ML estimates for $\boldsymbol{\beta}$ and $\boldsymbol{\omega}$ as follows: $\hat{\boldsymbol{\beta}}(\rho,\theta) =\left(\tilde{\textbf{X}}_o^\top\textbf{V}_{oo}^{-1}\tilde{\textbf{X}}_o\right)^{-1}\tilde{\textbf{X}}_o^\top\textbf{V}_{oo}^{-1}\textbf{y}_o~~~~ \text{and} ~~~~    \hat{\omega}(\rho,\theta)=\frac{\textbf{r}_{o}^\top\textbf{V}_{oo}^{-1}\textbf{r}_{o}}{n_o}$, where $\tilde{\textbf{X}}_o=\textbf{X}_o$ for H-SEM, and $\tilde{\textbf{X}}_o=\textbf{A}^{-1}\textbf{X}_o$ for H-SAM. By substituting $\hat{\boldsymbol{\beta}}(\rho,\theta)$, and $\hat{\omega}(\rho,\theta)$ in the log-likelihood in Equation~\eqref{eq:maginallik_me}, the concentrated marginal log-likelihood $L_c$ has the form: 
\begin{equation}
    \label{eq:con_log_lik_ME_miss}
    L_c(\theta,\rho)=c-\frac{n_o}{2}\textrm{log}(\hat{\omega}(\rho,\theta))-\frac{1}{2}\textrm{log}|\textbf{V}_{oo}|,
\end{equation} 
\noindent where $c=-\frac{n_o}{2}\textrm{log}(2\pi)-\frac{n_o}{2}$ is a constant. To obtain the ML estimates for $\theta$ and $\rho$, we maximise the concentrated marginal log-likelihood in Equation~\eqref{eq:con_log_lik_ME_miss} using \texttt{optim()} function in R. 

\subsection{Computational aspects of Marginal ML method}
\label{sec:compaccept}

The numerical optimisation of the concentrated marginal log-likelihood $L_c$, defined in Equation~\ref{eq:con_log_lik_ME_miss}, requires repeated evaluation of $L_c$ for different values of $\rho$ and $\theta$. Therefore, efficient computation of $L_c$ is critical to reduce the overall computational cost of the proposed MML algorithm. Below, we present two computational approaches for evaluating $L_c$.

Efficient computation of the matrix $\mathbf{V}_{oo}$ plays a key role in evaluating $L_c$. Our first approach, referred to as the \textit{direct} approach, involves directly extracting the sub-matrix $\mathbf{V}_{oo}$ from the larger matrix $\mathbf{V}$ as 
\begin{equation}
    \label{eq:v_oo}
    \textbf{V}_{oo}=[(\textbf{I}_n+\theta (\textbf{A}^\top\textbf{A})^{-1})]_{oo}.
\end{equation}
After computing $\mathbf{V}_{oo}$, the logarithms of its determinant and its inverse are calculated, and subsequently, $L_c$ is evaluated.

In the direct computational approach, the matrix $\textbf{V}$ is explicitly calculated. However, when evaluating $L_c$ in Equation~\eqref{eq:con_log_lik_ME_miss} and the closed-form ML estimators of $\boldsymbol{\beta}$ and $\omega$, only certain terms involving $\textbf{V}_{oo}$ are required. Specifically, we need to compute the following terms: (i) $\textbf{r}_o^\top\textbf{V}^{-1}_{oo}\textbf{r}_o$, (ii) $\tilde{\textbf{X}}_o^\top\textbf{V}_{oo}^{-1}\tilde{\textbf{X}}_o$, (iii) $\tilde{\textbf{X}}_o^\top\textbf{V}_{oo}^{-1}\textbf{y}_o$, and (iv) $\textrm{log}|\textbf{V}_{oo}|$. Directly computing these terms by first explicitly computing $\textbf{V}_{oo}$ can be computationally challenging, as $\textbf{V}_{oo}$ is a dense matrix. In our second computational approach, we reformulate these terms, along with $\mathbf{V}_{oo}$, to enable more efficient computation using sparse matrix operations.

In the second approach, referred to as the \textit{parameterisation}  approach, the sub-matrix $\mathbf{V}_{oo}$ is first extracted using an additional sparse matrix, $\textbf{B}_o$ from $\textbf{V}$ given in Table~\ref{tbl:properties}. This approach leverages the sparsity of $\textbf{W}$ and $\textbf{B}_o$ to simplify the calculation of  $\textbf{V}_{oo}$. We define the sparse matrix $\textbf{B}_o$ as $\textbf{B}_o=[\textbf{I}_{o}|\textbf{0}]$, where $\textbf{I}_{o}$ is the $n_o \times n_o$ identity matrix and \(\boldsymbol{0}\) is the $n_o \times n_u$ zero matrix. Now $\textbf{V}_{oo} = \textbf{B}_o \textbf{V} \textbf{B}_o^{\top}$. Then after some further simplifications, $\textbf{V}_{oo}$ can be reformulated as:
\begin{equation}
    \label{eq:V_oo1}
\begin{split}
    \textbf{V}_{oo}& = \textbf{B}_o \textbf{V} \textbf{B}_o^{\top}\\
    &=\textbf{B}_o (\textbf{I}_n+\theta (\textbf{A}^\top\textbf{A})^{-1}) \textbf{B}_o^{\top}\\
    &=\textbf{I}_o+\theta \textbf{B}_o  (\textbf{A}^\top\textbf{A})^{-1}\textbf{B}_o^{\top}.
\end{split}
\end{equation}
%\noindent where $\textbf{I}_o$, is the identity matrix of size $n_o$.

\begin{comment}
    \noindent By performing further simplifications, we obtain:
\begin{equation}
    \label{eq:re_expressterm1_}
\textbf{V}_{oo}={\textbf{C}^\top_o}(\textbf{A}\textbf{A}^\top+\theta\textbf{I})\textbf{C}_o,
\end{equation}
\noindent where $(\textbf{A}^\top)^{-1}\textbf{B}_{o}^{\top}=\textbf{C}_o$.
\end{comment}

\noindent First, the log determinant ($\log|\textbf{V}_{oo}|$) is computed. Using the matrix determinant lemma~\citep{DING20071223}, the logarithm of the determinant of $\textbf{V}_{oo}$ in Equation~\ref{eq:V_oo1} can be obtained as
\begin{equation}
    \label{eq:V_oodet}
    \text{log}|\textbf{V}_{oo}|=  \text{log}|\textbf{A}^\top\textbf{A}+\theta\textbf{B}_o^{\top}\textbf{B}_o|-\text{log}|\textbf{A}^\top\textbf{A}|,
\end{equation}

\noindent and the proof is given in Section~\ref{sec:sup_log_det} of the online supplement.

%The determinant of $\textbf{A}^\top\textbf{A} + \theta\textbf{B}_o^{\top}\textbf{B}_o$ can be efficiently obtained using its already computed Cholesky factors. Similarly, the determinant of $\textbf{A}^\top\textbf{A}$ can be efficiently computed using the Cholesky factors of $\textbf{A}^\top\textbf{A}$.

%The determinant of $\textbf{A}^\top\textbf{A}$ can be computed, fist computing choleskey factors of $\textbf{A}^\top\textbf{A}$, efficiently by using the \texttt{Cholesky()} function from the Matrix package~\citep{Matrix}. The determinant of $\textbf{A}^\top\textbf{A}+\theta\textbf{B}_o^{\top}\textbf{B}_o$, can be computed in a similar manner using its choleskey factor. However, using \texttt{updown()} function, the choleskey factor of $\textbf{A}^\top\textbf{A}+\theta\textbf{B}_o^{\top}\textbf{B}_o$ using the choleskey update can be computed with less computational effort compared to direct calculating the Choleskey factors. 

The determinant of $\mathbf{A}^\top \mathbf{A}$ can be efficiently computed by obtaining the Cholesky factors using the \texttt{Cholesky()} function from the Matrix package~\citep{Matrix}. Similarly, the determinant of $\mathbf{A}^\top \mathbf{A} + \theta \mathbf{B}_o^{\top} \mathbf{B}_o$ can be computed using its Cholesky factors. However, by employing the \texttt{updown()} function from the Matrix package, the Cholesky factors of $\mathbf{A}^\top \mathbf{A} + \theta \mathbf{B}_o^{\top} \mathbf{B}_o$ can be computed using the Cholesky factors of $\mathbf{A}^\top \mathbf{A}$ with less computational effort through the sparse Cholesky update~\citep{doi:10.1137/S0895479899357346}, compared to directly calculating the Cholesky factors of $\mathbf{A}^\top \mathbf{A} + \theta \mathbf{B}_o^{\top} \mathbf{B}_o$.

By applying the Woodbury matrix identity \citep[p. 427]{harville1997matrix} to the matrix $\mathbf{V}_{oo}$ in Equation~\eqref{eq:V_oo1}, we obtain $\mathbf{V}_{oo}^{-1}$ as follows:
\begin{equation}
    \label{eq:V_ooinv}
    \textbf{V}_{oo}^{-1}=\textbf{I}_o-\theta \textbf{B}_o  (\textbf{A}^\top\textbf{A}+\theta\textbf{B}_o^{\top}\textbf{B}_o)^{-1} \textbf{B}_o^{\top}.
\end{equation}
By substituting $ \textbf{V}_{oo}^{-1}$ in Equation~\eqref{eq:V_ooinv}, the required terms in $L_c$ are given by
\begin{equation}
    \label{eq:rotV_ooinvro}
    \textbf{r}_o^{\top}\textbf{V}_{oo}^{-1}  \textbf{r}_o=\textbf{r}_o^{\top}\textbf{r}_o-\theta \textbf{r}_o^{\top} \textbf{B}_o  (\textbf{A}^\top\textbf{A}+\theta\textbf{B}_o^{\top}\textbf{B}_o)^{-1} \textbf{B}_o^{\top}\textbf{r}_o,
\end{equation}

\begin{equation}
    \label{eq:xotV_ooinvxo}
    \tilde{\textbf{X}}_o^\top
    \textbf{V}_{oo}^{-1}\tilde{\textbf{X}}_o=\tilde{\textbf{X}}_o^\top\tilde{\textbf{X}}_o-\theta \tilde{\textbf{X}}_o^\top \textbf{B}_o  (\textbf{A}^\top\textbf{A}+\theta\textbf{B}_o^{\top}\textbf{B}_o)^{-1} \textbf{B}_o^{\top}\tilde{\textbf{X}}_o,
\end{equation}

\begin{equation}
    \label{eq:xotV_ooinvyo}
    \tilde{\textbf{X}}_o^\top\textbf{V}_{oo}^{-1}\textbf{y}_o= \tilde{\textbf{X}}_o^\top\textbf{y}_o-\theta \tilde{\textbf{X}}_o^\top \textbf{B}_o  (\textbf{A}^\top\textbf{A}+\theta\textbf{B}_o^{\top}\textbf{B}_o)^{-1} \textbf{B}_o^{\top}\textbf{y}_o.
\end{equation}

%for the required terms in $L_c$ evaluation, we get

%The matrix inverse $(\textbf{A}^\top\textbf{A}+\theta\textbf{B}_o^{\top}\textbf{B}_o)^{-1}$ in expressions given in Equations~\eqref{eq:rotV_ooinvro}, ~\ref{eq:xotV_ooinvxo}, and~\eqref{eq:xotV_ooinvyo} do not need to be calculated explicitly. Instead, first we calculate the choleskey factor of $\textbf{A}^\top\textbf{A}+\theta\textbf{B}_o^{\top}\textbf{B}_o$, and using it we use  forward and backward substitution to obtain these e terms. For instance to calculate the term $ (\textbf{A}^\top\textbf{A}+\theta\textbf{B}_o^{\top}\textbf{B}_o)^{-1} \textbf{B}_o^{\top}\textbf{r}_o$ in Equation~\ref{eq:rotV_ooinvro}, we solve the system  $ (\textbf{A}^\top\textbf{A}+\theta\textbf{B}_o^{\top}\textbf{B}_o)\textbf{a}=(\textbf{B}_o^{\top}\textbf{r}_o)$ for $\textbf{a}$ using forward and backward substitution. Given recalculated choleskey factors of $\textbf{A}^\top\textbf{A}+\theta\textbf{B}_o^{\top}\textbf{B}_o$, the \texttt{solve}() function from the Matrix package is used to implement forward and backward substitution.
%we can solve the systems 

\noindent The matrix inverse $(\textbf{A}^\top\textbf{A} + \theta\textbf{B}_o^{\top}\textbf{B}_o)^{-1}$, appear in Equations \eqref{eq:rotV_ooinvro}, \eqref{eq:xotV_ooinvxo}, and \eqref{eq:xotV_ooinvyo}, does not need to be computed explicitly. Instead, using the already computed Cholesky factors of $\textbf{A}^\top\textbf{A} + \theta\textbf{B}_o^{\top}\textbf{B}_o$, we perform forward and backward substitution to compute the required terms. For example, to compute the term $(\textbf{A}^\top\textbf{A} + \theta\textbf{B}_o^{\top}\textbf{B}_o)^{-1} \textbf{B}_o^{\top} \textbf{r}_o$ in Equation \eqref{eq:rotV_ooinvro}, we solve the system $(\textbf{A}^\top\textbf{A} + \theta\textbf{B}_o^{\top}\textbf{B}_o)\textbf{x} = \textbf{B}_o^{\top}\textbf{r}_o
$ for $\textbf{x}$ (where $\textbf{B}_o^{\top}\textbf{r}_o$ is a vector), using forward and backward substitution. Given the precomputed Cholesky factors of $\textbf{A}^\top\textbf{A} + \theta\textbf{B}_o^{\top}\textbf{B}_o$, the \texttt{solve()} function from the Matrix package is used to perform these operations efficiently. 
In the subsequent sections, we refer to the marginal ML methods as MML methods. The MML algorithm that employs the parameterisation computational approach is denoted as the MML-P method, whereas the implementation that utilises the direct computational approach is referred to as the MML-D method.

{The computational complexity of the MML-D method depends on matrix inversion of $\mathbf{A}^\top \mathbf{A}$ and the computation of the determinant and the inverse of $\mathbf{V}_{oo}$. Through a simulation study, we found that the complexity of inverting $\mathbf{A}^\top \mathbf{A}$ is approximately $O(n^2)$. In the simulation, we defined $\mathbf{W}$ on an $\sqrt{n} \times \sqrt{n}$ regular grid with a Rook local neighborhood (see Section~\ref{njanasec:sup_rook} of the online supplement). Both the determinant and inverse calculations of $\mathbf{V}_{oo}$ contribute a computational complexity of $O(n_{o}^3)$, as $\mathbf{V}_{oo}$ is a dense matrix. Thus, the overall computational complexity of the MML-D method is $\text{max}(O(n^2), O(n_o^3))$.}

{
In contrast, the computational complexity of the MML-P method primarily depends on computing the Cholesky factors of $\mathbf{A}^\top \mathbf{A}$. It can be demonstrated that $\mathbf{A}^\top \mathbf{A}$ exhibits a local neighborhood structure, provided that $\mathbf{W}$ possesses a local neighborhood structure; see Section~\ref{njanasec:sup_rook} of the online supplement for details. As shown by \citet{rue2005gaussian}, performing Cholesky factorisation on a sparse matrix with a local neighborhood constructed on a $\sqrt{n} \times \sqrt{n}$ regular grid has a computational complexity of $O(n^{3/2})$. Therefore, the Cholesky factorisation of $\mathbf{A}^\top \mathbf{A}$ also has a computational complexity of $O(n^{3/2})$, which leads to an overall computational complexity of the MML-P method of $O(n^{3/2})$.
}

%%%%%%%%%%%%%%%

%The direct approach gives an overall computational complexity of $O(n^2)$, where $n$ is the total number of observations; see Section~\ref{sec:supp_complexities} of the online supplement for further details.

%Since $\textbf{A}^\top\textbf{A} + \theta\textbf{B}_o^{\top}\textbf{B}_o$ is sparse for sparse $\textbf{W}$ matrices, calculating its Choleskey factor has computational complexity of $O(n^{3/2})$, typically. 

We strongly recommend using the MML-P method, not only for its lower computational complexity compared to the MML-D method but also for two additional important reasons. First, the parameterisation approach used in the MML-P method avoids the direct inversion of a sparse $n \times n$ matrix, which is required by the direct computational approach in the MML-D method. This direct inversion has a higher computational complexity of $O(n^2)$ and is less numerically stable. Second, although $\textbf{A}^\top\textbf{A}$ is sparse, its inverse is a dense matrix. As $n$ increases, the size of $(\textbf{A}^\top\textbf{A})^{-1}$ becomes large, making it impractical to store. In our simulation studies, attempts to store $(\textbf{A}^\top\textbf{A})^{-1}$ on the National Institute for Applied Statistics Research Australia High Performance Computer cluster\footnote{https://hpc.niasra.uow.edu.au/} failed when $n$ exceeded approximately $32,500$. By contrast, the parameterisation approach enabled us to compute the required terms even for larger values of $n=1,000,000$. Further, as shown in our simulation study in Section~\ref{sec:simulation}, and Table~\ref{tab:Lucas_SAM_SEM_10_obs} from our real-world analysis in Section~\ref{sec:example}, both MML-D and MML-P produce identical results, as expected. However, fitting H-SAR models including both parameter estimation and standard error calculation is substantially faster with MML-P. %\textcolor{red}{do we need to present time consumption to D and P for simulations as well?}

\section{Simulation study}
\label{sec:simulation}

This section discusses the accuracy of the MML-P method for different missing data percentages using simulated data. Additionally, We compare computation time consumption by MML-P and MML-D methods.

%In our simulation study, we utilise sparse weight matrices ($\mathbf{W}$), implemented using the R package \textit{Matrix}~\citep{matrixpkg}. We now explain the grid and neighbourhood structure used in constructing the spatial weight matrix $\textbf{W}$. Consider a regular grid of size $\sqrt{n} \times \sqrt{n}$, with $n$ observations. 
%We define neighbours based on the Rook neighbourhood structure~\citep{moura2020esda} as shown in Figure~\ref{fig:rook} in Section \ref{sec:supp_complexities} of the online supplement, where O is the unit of interest and, N's are neighbours of O. It is clear that any given unit can have $2$ to $4$ neighbours. 
%The rows of matrix $\textbf{W}$ (before row normalization) that correspond to Rook neighbourhoods can accommodate a maximum of 4 non-zero elements. The neighbourhood structure of $\textbf{W}$ is often called the W-neighbourhood in spatial econometrics literature~\citep{mukherjee2014spatially,suesse2018estimation}. 
%The details of the simulation study are given in Section~\ref{sec:supp_complexities} of the online supplement. 

%\subsection{Comparing Estimation Methods \label{sec:comparingestimation}}

%\textcolor{red}{Am i going to one MML method used or}

{In practice, we often encounter situations where the number of unobserved or missing units ($n_u$) is significantly larger than the number of observed units ($n_o$). For instance, in the property market, house prices can be modeled using the H-SAR models with missing data. Here, the prices of sold houses represent the observed responses, while the prices of unsold houses are treated as missing. Since only a small percentage of properties are sold each year, or even over a decade, $n_o$ is relatively small compared to the total number of properties ($n$), making $n_u$ much larger in comparison. Nevertheless, information on explanatory variables, such as house characteristics and location data for all properties, both sold and unsold are typically available from assessors~\citep{lesage2004models}.} 

When estimating SAR models with missing response data, researchers often overlook the missingness issue. In such cases, the SAR model employs a spatial weight matrix based only on the location data of the observed units. The likelihood constructed in this manner is referred to as the observed likelihood, and the estimates obtained from maximising the observed log-likelihood are termed \textit{observed maximum likelihood} (OML) estimates in the following sections of this paper. However, the observed likelihood does not reflect the marginal likelihood of the observed response variable, $\textbf{y}_o$, as it is based solely on the covariate matrix for the observed data, $\textbf{X}_o$, and the spatial weight matrix for the observed units, $\textbf{W}_{oo}$. This mis-specification results in biased and inconsistent parameter estimates~\citep{wang2013estimation,benedetti2020spatial}.
In contrast, correctly specifying the marginal distribution of $\textbf{y}_o$ requires incorporating the full covariate matrix, $\textbf{X}$, and the complete spatial weight matrix, $\textbf{W}$, which account for all $n$ spatial locations. This is the approach we propose in our MML method. To evaluate the performance of our MML method, we conducted a simulation study comparing its estimates to OML estimates.

To generate synthetic data, we set $\rho=0.8$, $\sigma^2_{\boldsymbol{\epsilon}}=2$, and $\sigma^2_{\textbf{e}}=1$ for both the H-SEM and H-SAM. The regression parameters are fixed at ${\beta}_0=1$ and ${\beta}_1=5$. The covariate is drawn from the standard normal distribution. The spatial weight matrix is generated based on a $71\times 71$ grid, using the Rook neighbourhood~\citep{moura2020esda}; see Section \ref{njanasec:sup_rook} of the online supplement for further details. In total, we simulate $250$ data sets, each with $ n=5041$ units. We use the R \texttt{optim}() function to numerically maximise the concentrated marginal log-likelihood function in Equation~\eqref{eq:con_log_lik_ME_miss} with respect to parameters $ \theta $ and $ \rho $. We set $10^{-8}$ as the convergence threshold for the \texttt{optim}() function.

%Table~\ref{tab:HSAM_HSEM_sim} presents the mean estimates and mean squared errors (MSEs) of $\rho$, $\sigma^2_{\boldsymbol{\epsilon}}$, and $\sigma^2_\textbf{y}$ for the H-SEM and H-SAM with $10\%$ and $50\%$ observed data from OML and MML . The tables include the average computational time. For both the H-SEM and H-SAM, regardless of the percentage of observed, the MML specifications consistently yield estimates closer to the true values with lower MSEs compared to OML estimates for all parameters. Section~\ref{sec:ext.simulation} of the online supplement provides further results for this study. These results include mean estimates, MSEs, mean standard errors of estimates, and the parameter coverages for all model parameters, including fixed effects ($\boldsymbol{\beta}$), across various levels of missing data percentages. %The analysis is based on 2500 datasets, each comprising 625 units.

%To compute the $\textbf{V}_{oo}$ matrix in MML, we employed the parameterisation computational approach.

Table~\ref{tab:HSAM_HSEM_sim} presents the mean parameter estimates and mean squared errors (MSEs) for the parameters $\rho$, $\sigma^2_{\boldsymbol{\epsilon}}$, and $\sigma^2_\textbf{e}$ in the H-SEM and H-SAM, with $90\%$ and $50\%$ missing data, estimated using the OML and MML-P methods across $250$ simulated datasets. For both models, the MML-P method consistently yields estimates that are closer to the true parameter values and achieves lower MSEs than the OML method, regardless of the percentage of missing data. Further results, including mean standard errors and coverage rates for all model parameters (including fixed effects $\boldsymbol{\beta}$), are provided in Tables~\ref{tab:sup_estimates_10per_obs_SEM} to \ref{tab:sup_estimates_50per_obs_SAM} in Section~\ref{sec:ext.simulation} of the online supplement. These additional results further demonstrate that the MML-P method offers better coverage and smaller mean standard errors compared to the OML method.

Both MML-D and MML-P methods yield identical results, but fitting H-SAR models (including both parameter estimation and standard error calculation) is faster with MML-P. In the simulation study presented in Table~\ref{tab:HSAM_HSEM_sim}, when $90\%$ of the data is missing, the average computation time for fitting the H-SEM is 671.9590  seconds using MML-D, compared to 12.3905 seconds with MML-P. Similarly, for the H-SAM, MML-D takes 675.9505 seconds, while MML-P takes 16.4584 seconds. For datasets with 50\% missing data, H-SEM fitting using the MML-D method took an average of 1293.76 seconds, whereas the MML-P method required only 10.02 seconds. Similarly, for H-SAM fitting with 50\% missing data, the MML-D method took an average of 1396.39 seconds, while the MML-P method required just 13.97 seconds.

%1.0689   0.7011 667.4944

\begin{table}[H]
\centering
\caption{Mean parameter estimates (with mean squared errors (MSEs) in the brackets), computed from $250$ simulated datasets, each with $n= 5,041$ units, for $\rho$, $\sigma^2_{\epsilon}$, and $\sigma^2_{\textbf{e}}$ for H-SEM and H-SAM with $90\%$ and $50\%$ missing data obtained using the OML and MML-P algorithms. The true parameter values are $\rho=0.8$, $\sigma^2_{\epsilon}=2$, and $\sigma^2_{\textbf{e}}=1$.}
\label{tab:HSAM_HSEM_sim}
\begin{tabular}{l@{\hspace{2pt}}r@{\hspace{2pt}}r@{\hspace{2pt}}r@{\hspace{2pt}}r@{\hspace{2pt}}r@{\hspace{2pt}}r@{\hspace{2pt}}r@{\hspace{2pt}}r@{\hspace{2pt}}}
\toprule
 & & \multicolumn{3}{c}{OML} &\multicolumn{3}{c}{MML-P} \\  \cmidrule(lr){3-5} \cmidrule(lr){6-8} 
                      & & \multicolumn{1}{c}{$\rho$}   & \multicolumn{1}{c}{$\sigma^2_{\epsilon}$} & \multicolumn{1}{c}{$\sigma^2_{\textbf{e}}$} & \multicolumn{1}{c}{$\rho$}  & \multicolumn{1}{c}{$\sigma^2_{\epsilon}$} & \multicolumn{1}{c}{ $\sigma^2_{\textbf{e}}$}\\ 
                      \cmidrule(lr){1-2} \cmidrule(lr){3-5} \cmidrule(lr){6-8} 
\multirow{3}{*}{H-SEM} & $90\%$   & \makecell{ 0.2557\\ (0.3265)}  & \makecell{1.0016 \\(2.5969)} & \makecell{ 3.1122 \\ (6.2679)} &  \makecell{ 0.7880\\ (0.0094)} & \makecell{ 1.9189\\(0.6545)} & \makecell{1.1157 \\ (0.4689)}  \\
                     & $50\%$   &  \makecell{0.6109 \\ (0.0413)}& \makecell{2.3856 \\ (0.2385)} & \makecell{0.9737 \\(0.1336)} & \makecell{ 0.7949\\ (0.0012)} & \makecell{1.9745\\ (0.0567)} & \makecell{1.0350 \\ ( 0.0548)} \\ 
\cmidrule(lr){1-2} \cmidrule(lr){3-5} \cmidrule(lr){6-8} 
\multirow{3}{*}{H-SAM} & $90\%$   & \makecell{0.2434 \\ (0.3107)}    & \makecell{0.0020 \\(3.9921)} & \makecell{ 19.0154 \\ (327.01)} & \makecell{0.8003 \\ (0.0001)} & \makecell{2.0111 \\ (0.2587)} & \makecell{ 0.9748 \\(0.0674)} \\
                     & $50\%$   & \makecell{0.4497\\(0.1229)}   &\makecell{0.0374 \\(3.8650)} &  \makecell{10.4758 \\(89.9968)}&   \makecell{0.7997\\(0.0001)}& \makecell{2.0059\\(0.0247)} & \makecell{0.9995\\ (0.0115)}  \\
\bottomrule
\end{tabular}
\end{table}

\section{Real Data Application}

\label{sec:example}

This section applies the estimation methods to a real dataset. 
The dataset consists of house prices from Lucas County, Ohio, USA, including $25,357$ observations of single-family homes sold between 1993 and 1998. A detailed description of the dataset is available through the Spatial Econometrics toolbox for Matlab\footnote{The dataset can be accessed at \href{http://www.spatial-econometrics.com/html/jplv7.zip}{http://www.spatial-econometrics.com/html/jplv7.zip}}~\textcolor{red}. The dataset is also included in the R package spData ~\citep{spData}.
We use the natural logarithm of housing prices (\(\ln(\text{price})\)) as the dependent variable and include several independent variables, including different powers of house age ($\text{age}, \text{age}^2$, and $\text{age}^3$), the logarithm of the lot size in square feet (\(\ln(\text{lotsize})\)), the number of rooms (\text{rooms}), the logarithm of the total living area in square feet (\text{LTA}), the number of bedrooms, and a binary indicator for each year from 1993 to 1998 (\text{syear}) to represent the year of the house sale. The same row-normalised sparse weight matrix $\textbf{W}$, as introduced by ~\cite{bivand2010comparing}, is used. 

In this section, we compare the estimates from the OML and MML methods with those from the full maximum likelihood (FML) estimation method. The FML method refers to the ML estimation of H-SEM and H-SAM using the entire dataset, which contains no missing values. Here, we use a dataset without missing values so that it is possible to estimate the parameters using the FML method. The estimates from FML are used as the ground truth for comparing the OML and MML methods. To better simulate real-world scenarios, we introduce moderate to high levels of missing data in the response variable. Specifically, we create two datasets: one with $50\%$ missing responses and another with $90\%$ missing responses, obtained from the complete dataset. We set the convergence threshold for the \texttt{optim()} function to $10^{-8}$.

%Table~\ref{tab:Lucas_SAM_SEM_10_obs} presents the estimates and standard errors for $\rho$, $\sigma^2_{\epsilon}$, and $\sigma^2_{\textbf{e}}$ obtained using the FML, OML, and MML methods for the H-SEM and H-SAM models with $10\%$ of the observed data. The table also includes the estimation time and the time required to compute the standard errors of the parameters. The results are consistent with our findings provided in Section~\ref{sub.sec:complexity}. The estimates obtained from MML are closer to the estimates obtained from the FML than those of the OML, which relies solely on observed data locations; see Table ~\ref{tab:sup.Lucas_SEM_25} in Section ~\ref{sec:sup.real} of the online supplement for the analysis of data set with $50\%$ of observed data, and which results consist with the results from data set with $10\%$ observed. Further, Tables~\ref{tab:sup.Lucas_SEM_25}, and \ref{} of the online supplement contain estimates and standard errors of fixed effects ($\boldsymbol{\beta}$) for both H-SEM and H-SAM, with $10\%$, and $50\%$ of observed data.

Table~\ref{tab:Lucas_SAM_SEM_10_obs} presents the estimates and standard errors for parameters $\rho$, $\sigma^2_{\epsilon}$, and $\sigma^2_{\textbf{e}}$ obtained from the FML, OML, and MML methods for the H-SEM and H-SAM, based on the dataset with $90\%$ missing data. The table also reports the computation times for both parameter estimation and the calculation of standard errors. The results are consistent with our findings in Section~\ref{sec:simulation}, demonstrating that the MML estimates (for both MML-D and MML-P methods) align more closely with the FML estimates compared to the OML estimates, which rely only on observed data locations. As anticipated, the time required for both parameter estimation and standard error computation is substantially lower for both models when using the proposed MML-P method compared to MML-D.

Tables~\ref{tab:sup_Lucas_SEM_50}, and~\ref{tab:sup_Lucas_SAM_50} in Section~\ref{sec:sup.real} of the online supplement show the results for the dataset with $50\%$ missing data. The results are consistent with those from the dataset with $90\%$ missing data.
In addition, Tables~\ref{tab:sup_Lucas_SEM_10} to~\ref{tab:sup_Lucas_SAM_50} in Section \ref{sec:sup.real} of the online supplement provide estimates and standard errors for all model parameters including fixed effects ($\boldsymbol{\beta}$) for both the H-SEM and H-SAM, accounting for both $50\%$ and $90\%$ missing data.

\begin{table}[H]
\captionsetup{justification=centering}
\caption{Parameter estimates (est) with their standard errors (se), and computation time in seconds (ct) for fitting H-SAM and H-SEM using FML, OML, MML-D, and MML-P for the dataset with $90\%$ missing data.}
\label{tab:Lucas_SAM_SEM_10_obs}
\centering
\begin{tabular}{cccccccccc}
\hline
\multirow{2}{*}{Model} &\multirow{2}{*}{} & \multicolumn{2}{c}{FML}  & \multicolumn{2}{c}{OML} & \multicolumn{2}{c}{\makecell{MML-D}} & \multicolumn{2}{c}{\makecell{MML-P}}\\ \cline{3-10}  
                         &  &              est               &   se       & est         & se      & est & se    & est & se    \\
                         \hline

\textcolor{black}{\multirow{5}{*}{H-SEM}} &$\rho$ & 0.9866 & 0.0002 &  0.6866& 0.0012& 0.9936& 0.0006 & 0.9936& 0.0006   \\
&$\sigma^2_{\textbf{e}}$ & 0.0004 & 0.0001 & 0.1643 & 0.0030 & 0.0001& 0.0002 & 0.0001& 0.0002   \\
&$\sigma^2_{\epsilon}$ & 0.0685 & 0.0007 & 0.0001 & 0.0081 & 0.0837 & 0.0035 & 0.0837 & 0.0035 
\\ \cline{2-10}  
&                   ct &  \multicolumn{2}{c}{ 20.098} & \multicolumn{2}{c}{  1.39} & \multicolumn{2}{c}{2792.409} & \multicolumn{2}{c}{38.65}  \\
                   \hline 
                   \hline

\multirow{5}{*}{H-SAM} &$\rho$ & 0.6727 & 0.0001 & 0.0027&0.0311 & 0.6046 & 0.0201   & 0.6046 & 0.0201   \\
&$\sigma^2_{\textbf{e}}$ & 0.0399 & 0.0008 &0.0882 &0.0114 &  0.0682 & 0.0070 &0.0682 & 0.0070   \\
&$\sigma^2_{\epsilon}$ & 0.042 & 0.0009 &  0.0882 &0.0234 & 0.0203& 0.0080  & 0.0203& 0.0080 
\\ \cline{2-10}  
&                  ct & \multicolumn{2}{c}{12.245} & \multicolumn{2}{c}{1.08} &\multicolumn{2}{c}{1196.3}  & \multicolumn{2}{c}{22.53}  \\
                   \hline 
\end{tabular}
\end{table}

To compare the accuracy of the OML and MML-P estimates, we computed the average mean squared errors (MSEs) of the estimated parameters for the H-SEM and H-SAM, as presented in Table~\ref{tab:MSE.real}. The estimates obtained from the FML are treated as the true parameter values. The average MSE for the MML-P estimates is calculated using the formula: $\text{MSE}_{\text{MML-P}} = \frac{1}{r+4} \sum_{i=1}^{r+4} (\hat{\theta}_{i} - \hat{\theta}_{i,\text{MML-P}})^2$, where $\hat{\theta}_{i}$ is the estimated value of the $i$-th parameter from the FML, $\hat{\theta}_{i,\text{MML-P}}$ is the corresponding parameter estimate from the MML-P method, and $r+4$ is the total number of parameters with $r$ the number of fixed effect parameters (excluding the intercept parameter). Similarly, the average MSE for the OML estimates is calculated in a similar manner.

%Table~\ref{} provides the MSEs for parameters from both the H-SEM and H-SAM models, based on datasets with 10\% and 50\% missing data.
%\textcolor{red}{need to change:}

\begin{table}[H]
  \centering
    \caption{Average mean squared error (MSE) of OML and MML-P estimates (relative to the FML estimates) for different missing data percentages.}
    \label{tab:MSE.real}
  \begin{tabular}{cllll}
    \hline
      & \multicolumn{2}{c}{H-SEM}  & \multicolumn{2}{c}{H-SAM} \\
      & \makecell{\textcolor{black}{$50\%$}} & \makecell{$90\%$ }& \makecell{\textcolor{black}{$50\%$}} & \makecell{$90\%$ } \\
          \hline
   OML &  0.7029 &  0.4838 &  1.2573 & 1.1248 \\

  MML-P &  0.0153 & 0.2556   &  0.0087 &  0.0172  \\
   \hline
  \end{tabular}
\end{table}
As expected, the average MSEs of parameter estimates obtained from the OML method are significantly higher than that of the MML-P method, regardless of the missing data percentages, for both models. These results highlight the importance of using the proposed MML-P method, particularly in real-world situations where high percentages of missing data are common. 

\section{Conclusion}

\label{sec:conclusion}
This article makes two key contributions. First, it introduces new H-SAR models that account for measurement errors and missing data in the response variable. Second, we propose the marginal maximum likelihood (MML) methods to accurately estimate the parameters of these H-SAR models. In particular, we propose the parameterisation method, that improves the computational efficiency of the MML method greatly when $n$
is large, as is often the case in real-world scenarios. 
Although this study focuses on estimating H-SAR models under the missing at random (MAR) mechanism, future research should extend to estimation under the missing not-at-random (MNAR) mechanism.

%(the number of missing units) 

\section*{Statements and Declarations}
Conflict of interest: The authors declare no potential or apparent conflict of interest in this article.

\begin{singlespace}
\bibliographystyle{apalike}
\bibliography{references.bib}
%\bibliography{main_final.bbl}
\end{singlespace}

\pagebreak

\renewcommand{\thefigure}{S\arabic{figure}} %This command modifies the numbering format for figures.

\renewcommand{\theequation}{S\arabic{equation}} %This command changes the numbering format for equations to "S1.1, S1.2, ...".

\renewcommand{\thetable}{S\arabic{table}}

\renewcommand{\thealgorithm}{S\arabic{algorithm}}

\renewcommand{\thesection} % setting sections
{S\arabic{section}}

%\onehalfspacing

\section*{Online Supplement for A Marginal Maximum Likelihood Approach for
Hierarchical Simultaneous Autoregressive Models with
Missing Data}

\setcounter{page}{1} % start page numbering from 1
\setcounter{section}{0} % start section numbering from 1

\setcounter{equation}{0} % start section numbering from 1
\setcounter{table}{0} % start section numbering from 1
\setcounter{figure}{0} % start section numbering from 1
\setcounter{algorithm}{0} % start section numbering from 1

We use the following notation in the supplement. Eq.~(1), Table~1,
and Figure~1, etc, refer to the main paper, while Eq.~(S1.1),
Table~S1.1, and Figure~S1.1, etc, refer to the supplement.

\section{Construction of spatial weight matrix for simulations}
\label{njanasec:sup_rook}

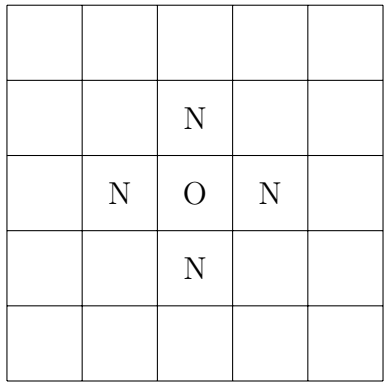
\begin{figure}[H]
  \centering
    \begin{tikzpicture}
      % Grid lines
      \draw (0,0) grid (5,5);
      % Add "an" in the fifth unit
      \node at (2.5,2.5) {O};
       \node at (2.5,1.5) {N};
        \node at (2.5,3.5) {N};
         \node at (1.5,2.5) {N};
          \node at (3.5,2.5) {N};
    \end{tikzpicture}
    \caption{The grid used for constructing $\textbf{W}$ based on the Rook neighbourhood (also referred to as the W-neighbourhood).}
    \label{fig:w}
  \label{fig:rook}
\end{figure}

We briefly explain the grid and neighbourhood structure used to construct the spatial weight matrix $\textbf{W}$, which is employed in the simulation studies presented in Section~\ref{sec:simulation} of the main document, as well as in Sections~\ref{sec:invAAT} and~\ref{sec:ext.simulation} of the online supplement. Consider a regular grid of size $\sqrt{n} \times \sqrt{n}$, with $n$ observations. 
We define neighbours based on the Rook neighbourhood criterion~\citep{moura2020esda} as shown in Figure~\ref{fig:rook}, where O is the unit of interest and, N's are neighbours of O. It is clear that any given unit can have $2$ to $4$ neighbours. 
The rows of matrix $\textbf{W}$ that correspond to Rook neighbourhoods can accommodate a maximum of 4 non-zero elements.  The neighbourhood structure of $\textbf{W}$ is often called the W-neighbourhood in spatial econometrics literature~\citep{mukherjee2014spatially,suesse2018estimation}. Moreover, since each row of $\textbf{W}$ contains at most 4 non-zero elements, regardless of $n$, the neighbourhood structure of $\textbf{W}$ is referred to as a local neighbourhood.

%Therefore, it can be concluded that the Rook neighborhood is a local neighborhood

%Figure~\ref{fig:rook} (b) explains the neighbors of a unit of interest in matrices $\textbf{W}\textbf{W}^\top$ and $\textbf{W}^\top\textbf{W}$, with a similar pattern observed for both cases. Similar to the W-neighborhood, it can be observed that both $\textbf{W}\textbf{W}^\top$ and $\textbf{W}^\top\textbf{W}$ have a maximum number of neighbors bounded by a constant. This suggests that both $\textbf{W}\textbf{W}^\top$ and $\textbf{W}^\top\textbf{W}$ represent local neighborhoods. 

%Figure~\ref{fig:rook} (b) explains the structure of $\textbf{W}\textbf{W}^\top$ and $\textbf{W}^\top\textbf{W}$, with a similar pattern observed for both cases. Specifically, in $\textbf{W}\textbf{W}^\top$ or $\textbf{W}^\top\textbf{W}$, two units $i$ and $j$ are  neighbours if they have a common neighbor, in the grid. Similar to the W-neighborhood, it can be observed that both $\textbf{W}\textbf{W}^\top$ and $\textbf{W}^\top\textbf{W}$ have a maximum number of neighbors bounded by a constant.  This suggests that both $\textbf{W}\textbf{W}^\top$ and $\textbf{W}^\top\textbf{W}$ represent local neighborhoods. 
It is worth noting that since $\textbf{W}$ represents a local neighborhood, $\textbf{A}^\top\textbf{A} = (\textbf{I} - \rho \textbf{W})^\top(\textbf{I} - \rho \textbf{W})$ also retains a local neighborhood structure.

\section{Computational complexity analysis of $(\textbf{A}^\top\textbf{A})^{-1}$}
\label{sec:supp_complexities}
\label{sec:invAAT}
This section approximates the empirical computational complexity of inverting $\textbf{A}^\top\textbf{A}$, which is employed in the implementation of the MML-D method using the direct computational approach described in Section~\ref{sec:compaccept} of the main document.

%The theoretical computational complexity of a matrix operation is established by analysing the structure of the matrix (or matrices) involved in the specific operation. To find the theoretical complexity of a specific matrix operation, we need to compute the number of scalar operations, known as flops (floating-point operations).  For instance, consider the multiplication of two $n \times n$ dense matrices. This operation comprises $n^3$ scalar multiplications and $n^2(n-1)$ scalar additions. In total, the computation involves $2n^3-n^2$ flops. As a consequence, the overall computational complexity is $O(n^3)$.

To address potential numerical issues that may arise when directly inverting $\textbf{A}^\top\textbf{A}$, we solve the system $\left(\textbf{A}^\top\textbf{A}\right)\textbf{X} = \textbf{I}$ for the matrix $\textbf{X}$, using the \texttt{solve()} function from the Matrix package~\citep{Matrix}. As detailed in Section~\ref{sec:compaccept} of the main document, the Cholesky factorisation of $\textbf{A}^\top\textbf{A}$ is employed to efficiently solve this system. Given that $\textbf{A}^\top\textbf{A}$ corresponds to a local neighbourhood structure, its Cholesky factorisation has a computational complexity of $O(n^{3/2})$. Let the Cholesky factor of $\textbf{A}^\top\textbf{A}$ be denoted as $\textbf{L}_{\textbf{A}^\top\textbf{A}}$. The system to be solved can be written as:
\begin{equation}
    \begin{split}
    \label{eq:solveAAT}
    \left(\textbf{A}^\top\textbf{A}\right)\textbf{X}&=\textbf{I},\\
           \left(\ {\textbf{L}}_{\textbf{A}^\top\textbf{A}}{({\textbf{L}}_{\textbf{A}^\top\textbf{A}})}^\top\right)\textbf{X}&=\textbf{I},\\
                   \ {\textbf{L}}_{\textbf{A}^\top\textbf{A}}\left({({\textbf{L}}_{\textbf{A}^\top\textbf{A}})}^\top\textbf{X}\right)&=\textbf{I}.\\
    \end{split}
\end{equation}

To solve the system in Equation~\eqref{eq:solveAAT}, a two-step approach is used. First, forward substitution is applied to the system ${\textbf{L}}_{\textbf{A}^\top\textbf{A}} \textbf{Z} = \textbf{I}$, yielding $\textbf{Z}$. Then, using $\textbf{Z}$, backward substitution is performed on ${\left({\textbf{L}}_{\textbf{A}^\top\textbf{A}}\right)}^\top \textbf{X} = \textbf{Z}$, resulting in $\textbf{X}$, the inverse of $\textbf{A}^\top\textbf{A}$.

 During the application of forward substitution to ${{\textbf{L}}_{\textbf{A}^\top\textbf{A}}}\textbf{Z}=\textbf{I}$ for obtaining $\textbf{Z}$, it is notable that ${{\textbf{L}}_{\textbf{A}^\top\textbf{A}}}$ forms a lower triangular matrix. Assuming all the lower triangular elements are non-zero, the computation of one row of the matrix $\textbf{Z}$ requires $n^2$ number of scalar operations (flops). As $\textbf{Z}$ possesses $n$ columns, the total number of flops needed to compute its all entries is  $n^3$. Similarly, the application of backward substitution to ${({\textbf{L}}_{\textbf{A}^\top\textbf{A}})}^\top\textbf{X}=\textbf{Z}$ for obtaining $\textbf{X}$ assuming all the upper triangular elements in $({\textbf{L}}_{\textbf{A}^\top\textbf{A}})^\top$ are non-zero, requires $n^3$ flops.

 However, it is essential to note that $\textbf{A}^\top\textbf{A}$ is a sparse matrix, which implies that its Cholesky factors are also sparse. As a result, a significant number of entries in the lower triangular matrix ${\textbf{L}}_{\textbf{A}^\top\textbf{A}}$ are zero. This sparsity reduces the number of flops required for both forward and backward substitutions, making it less than $n^3$. Consequently, the complexity of these operations is lower than $O(n^3)$. Deriving the theoretical computational complexity for forward and backward substitutions, in this case, is challenging; therefore, we conduct a simulation study to approximate the empirical computational complexity of these operations.
%(the scalar operations that determine the computational complexity of particular matrix operations) 

We now illustrate the simulation procedure for estimating the computational complexity of performing forward and backward substitution using the \texttt{solve()} function.
First, we construct the spatial weight matrix $\textbf{W}$ based on the Rook neighbourhood. Next, we derive $\textbf{A} = \textbf{I}_n - \rho\textbf{W}$ and compute $\textbf{A}^\top\textbf{A}$, followed by calculating the Cholesky factors ${\textbf{L}}_{\textbf{A}^\top\textbf{A}}$. We then perform forward and backward substitution repeatedly using ${\textbf{L}}_{\textbf{A}^\top\textbf{A}}$ and the \texttt{solve()} function for a fixed $n$. The average computation time is calculated from 5000 replications. This process is repeated for various values of $n$, and the resulting computation times are plotted against their respective $n$ values. Finally, the following regression line \begin{equation}
    \label{eq:computational.comp}
    t_i={b}\times n_i^\alpha,
\end{equation}
\noindent is fitted to estimate the computational complexity, where ${t}_i$ denotes the average computation time of performing forward and backward substitution when the total number of observations $n=n_i$, $b$ is a constant, and $\alpha \geq 0$ is the computational complexity. However, the estimation of the regression is done in the log scale; $\text{log}({t}_i)=\text{log}(b)+\alpha \times \text{log}({n}_i)$, using the least square method. Based on this simulation study, we estimate that the computational complexity for performing forward and backward substitution is $O(n^{2.057})\approx O(n^{2})$. 

In summary, to compute the inverse of $\textbf{A}^\top\textbf{A}$ we initially compute the Cholesky factors of $\textbf{A}^\top\textbf{A}$, which has a complexity of $O(n^{3/2})$. Subsequently, the application of forward and backward substitution, benefiting from the sparsity of $\textbf{A}^\top\textbf{A}$, involves a complexity of less than $O(n^3)$. Through the simulation study, we estimate that the empirical computational complexity of this step is approximately $O(n^{2})$. Consequently, the overall complexity of inverting $\textbf{A}^\top\textbf{A}$ is given by $\max(O(n^{3/2}), O(n^{2}))$, which simplifies to $O(n^{2})$.

\vspace{0.2cm}

\section{Extended simulation results}
\label{sec:ext.simulation}

%Table~\ref{tab:estimates_10per_obs_SEM} to~\ref{tab:estimates_50per_obs_SAM} present the mean estimates and mean squared errors of parameter estimates and their coverages obtained from $250$ simulated datasets (each with $n=5,041$ units) for the H-SEM and H-SAM. The coverage calculation is determined using $95\%$ Wald-type confidence intervals, employing the formula $\hat{\delta} \pm 1.96\sqrt{\widehat{\text{Var}}(\hat{\delta})}$, where $\delta \in \boldsymbol{\phi}$ and $\widehat{\text{Var}}(\hat{\delta})$ is the estimated variance of the parameter $\delta$. The estimated variances for the parameters are computed based on the formula given in Section~\ref{sec:info} of the online supplement.

Tables~\ref{tab:sup_estimates_10per_obs_SEM} to~\ref{tab:sup_estimates_50per_obs_SAM} show the mean estimates, mean standard errors, and coverage rates of parameter estimates obtained from 250 simulated datasets, each with $n=5,041$ units. These tables cover both the H-SEM and H-SAM across various percentages of missing data. Coverage rates are calculated using $95\%$ Wald-type confidence intervals, given by the formula $\hat{\delta} \pm 1.96\sqrt{\widehat{\text{Var}}(\hat{\delta})}$, where $\delta \in \boldsymbol{\phi}$ and $\widehat{\text{Var}}(\hat{\delta})$ represents the estimated variance of the parameter $\delta$. The estimated variances are computed using the formula provided in Section~\ref{sec:info} of the online supplement.

%\subsection{Missing percentage 10\%}

%\textcolor{red}{final 10 SEM and SAM}

\begin{table}[H]
\captionsetup{justification=centering}
\caption{H-SEM mean estimates (est.), mean standard error (se.), and coverage (cov.) over 250 simulated datasets of estimated parameters obtained using the OML and MML-P methods- $90\%$ missing data.}
\label{tab:sup_estimates_10per_obs_SEM}
\centering
\begin{tabular}{rrrrrrrr}
\toprule
\multirow{2}{*}{ } & \multirow{2}{*}{ \begin{turn}{90}
  \makecell{True \\ value}
\end{turn}} & \multicolumn{3}{c}{OML} & \multicolumn{3}{c}{MML-P} \\ 
\cmidrule(lr){3-5} \cmidrule(lr){6-8}
                           &                             & \multicolumn{1}{c}{est.}         &  \multicolumn{1}{c}{se.} & \multicolumn{1}{c}{cov.}        & \multicolumn{1}{c}{est.}         &  \multicolumn{1}{c}{se.} & \multicolumn{1}{c}{cov.}   \\ \midrule
$\beta_0$ & 1 &  0.9982 & 0.0954 & 0.9219 &  0.9989 & 0.1149 & 0.9766 \\
$\beta_1$ & 5 & 5.0273  & 0.0906 & 0.9453  &  5.0264 & 0.0894 &  0.9375  \\
$\rho$  & 0.8 & 0.2557 & 0.2958 &  0.3281 &  0.7880& 0.1536 &  0.8594   \\
$\sigma^2_{\boldsymbol{\epsilon}}$  & 2 & 1.0016 & 3.6828 & 0.6641 & 1.9189    & 1.3436 & 0.9375 \\
$\sigma^2_\textbf{e}$ & 1 & 3.1122 & 3.7556 & 0.5625&  1.1157  & 1.2624 & 0.8594 \\

\bottomrule
\end{tabular}
\end{table}

\begin{table}[H]
\captionsetup{justification=centering}
\caption{H-SAM mean estimates (est.), mean standard error (se.), and coverage (cov.) over 250 simulated datasets of estimated parameters obtained using the OML and MML-P methods- $90\%$ missing data.}
\label{tab:sup_estimates_10per_obs_SAM}
\centering
\begin{tabular}{rrrrrrrr}
\toprule
\multirow{2}{*}{ } & \multirow{2}{*}{ \begin{turn}{90}
  \makecell{True \\ value}
\end{turn}} & \multicolumn{3}{c}{OML} & \multicolumn{3}{c}{MML-P} \\ 
\cmidrule(lr){3-5} \cmidrule(lr){6-8}
                           &                             & \multicolumn{1}{c}{est.}         &  \multicolumn{1}{c}{se.} & \multicolumn{1}{c}{cov.}        & \multicolumn{1}{c}{est.}         &  \multicolumn{1}{c}{se.} & \multicolumn{1}{c}{cov.}      \\ \midrule
$\beta_0$ & 1 & 4.6018  &  0.1944 & 0.0000 & 0.9970  & 0.0224 &   0.6493 \\
$\beta_1$ & 5 &  6.2328 & 0.1949 & 0.0000& 4.9999  & 0.0624 &   0.8134 \\
$\rho$  & 0.8 &  0.2434  & 0.0332 &0.0000 & 0.8003 & 0.1600  &    0.9870\\
$\sigma^2_{\boldsymbol{\epsilon}}$  & 2 & 0.0020  &  6.5103 & 0.9732 &2.0111    &  1.2480 &    0.9825\\
$\sigma^2_\textbf{e}$ & 1   & 19.0154   &  6.3165 & 0.0746 & 0.9748 & 1.0835  &   0.9825 \\

\bottomrule
\end{tabular}
\end{table}

\begin{table}[H]
\captionsetup{justification=centering}
\caption{H-SEM mmean estimates (est.), mean standard error (se.), and coverage (cov.) over 250 simulated datasets of estimated parameters obtained using the OML and MML-P methods- $50\%$ missing data.}
\label{tab:sup_estimates_50per_obs_SEM}
\centering
\begin{tabular}{rrrrrrrr}
\toprule
\multirow{2}{*}{ } & \multirow{2}{*}{ \begin{turn}{90}
  \makecell{True \\ value}
\end{turn}} & \multicolumn{3}{c}{OML} & \multicolumn{3}{c}{MML-P} \\ 
\cmidrule(lr){3-5} \cmidrule(lr){6-8}
                           &                             & \multicolumn{1}{c}{est.}         &  \multicolumn{1}{c}{se.} & \multicolumn{1}{c}{cov.}        & \multicolumn{1}{c}{est.}         &  \multicolumn{1}{c}{se.} & \multicolumn{1}{c}{cov.}    \\ \midrule
$\beta_0$ & 1 &  1.0037  &  0.0781 & 0.8333 &  1.0027 &  0.0539  &  0.9867\\
$\beta_1$ & 5 & 4.9937  & 0.0387 & 0.9467 & 4.9935 &  0.0374 & 0.9533  \\
$\rho$  & 0.8 & 0.6109  & 0.0648 &  0.1933 &  0.7949   & 0.0361 & 0.9533   \\
$\sigma^2_{\boldsymbol{\epsilon}}$  & 2 &   2.3856  & 0.2795 & 0.6133 &   1.9745  & 0.2456 & 0.9533\\
$\sigma^2_\textbf{e}$ & 1 & 0.9737  & 0.3248 & 0.8600 & 1.0350  &  0.2369 & 0.9400  \\

\bottomrule
\end{tabular}
\end{table}

\begin{table}[H]
\captionsetup{justification=centering}
\caption{H-SAM mean estimates (est.), mean standard error (se.), and coverage (cov.) over 250 simulated datasets of estimated parameters obtained using the OML and MML-P methods- $50\%$ missing data.}
\label{tab:sup_estimates_50per_obs_SAM}
\centering
\begin{tabular}{rrrrrrrr}
\toprule
\multirow{2}{*}{ } & \multirow{2}{*}{ \begin{turn}{90}
  \makecell{True \\ value}
\end{turn}} & \multicolumn{3}{c}{OML} & \multicolumn{3}{c}{MML-P} \\ 
\cmidrule(lr){3-5} \cmidrule(lr){6-8}
                           &                             & \multicolumn{1}{c}{est.}         &  \multicolumn{1}{c}{se} & \multicolumn{1}{c}{cov.}        & \multicolumn{1}{c}{est.}         &  \multicolumn{1}{c}{se} & \multicolumn{1}{c}{cov.}       \\ \midrule
$\beta_0$ & 1 & 2.9032 & 0.0648 & 0.0000&1.0007  & 0.0141  &  0.8151  \\
$\beta_1$ & 5 & 5.6513  & 0.0648  &  0.0000 & 5.0009  & 0.0300 &0.8699   \\
$\rho$  & 0.8 & 0.4497 & 0.0100 & 0.0000 &0.7997   & 0.0361 & 0.9767  \\
$\sigma^2_{\boldsymbol{\epsilon}}$  & 2 & 0.0374  &  0.5297&0.0205 &  2.0059 & 0.2373 &  0.9732  \\
$\sigma^2_\textbf{e}$ & 1 & 10.4758  & 0.6581 & 0.0000&  0.9995  & 0.2236 & 0.9867 \\
\bottomrule
\end{tabular}
\end{table}

%In summary, our findings of the simulation study presented in Tables~\ref{tab:sup_estimates_10per_obs_SEM} to~\ref{tab:sup_estimates_50per_obs_SAM} indicate that estimates using the OML method yields inaccurate parameter estimates across both levels of missing data. In contrast, the MML method consistently provides highly accurate parameter estimates for both the H-SEM and H-SAM models, demonstrating superior coverage compared to OML.

In summary, the simulation study presented in Tables~\ref{tab:sup_estimates_10per_obs_SEM} to~\ref{tab:sup_estimates_50per_obs_SAM}  reveals that the OML method yields inaccurate parameter estimates across different missing data percentages. In contrast, the MML-P method consistently produces highly accurate parameter estimates for both the H-SEM and H-SAM and has significantly better coverage than OML.

\section{Extended Real data analysis results}
\label{sec:sup.real}

%   Tables~\ref{tab:sup_Lucas_SEM_10}, ~\ref{tab:sup_Lucas_SAM_10}, ~\ref{tab:sup_Lucas_SEM_50}, and ~\ref{tab:sup_Lucas_SAM_50}  displays estimates and standard errors for fixed effects ($\boldsymbol{\beta}$'s), $\rho$, $\sigma^2_{\epsilon}$, and $\sigma^2_{\textbf{e}}$ for the H-SEM with $90\%$ missing values,  H-SAM with $90\%$ missing values,  H-SEM with $50\%$ missing values,  H-SAM with $50\%$ missing values, respectively obtained using FDM, ODM, and MML methods. The table includes the computing time, the time taken to compute the standard errors of the parameter estimates.

Tables~\ref{tab:sup_Lucas_SEM_10}, ~\ref{tab:sup_Lucas_SAM_10}, ~\ref{tab:sup_Lucas_SEM_50}, and ~\ref{tab:sup_Lucas_SAM_50} present estimates and standard errors for fixed effects ($\boldsymbol{\beta}$), $\rho$, $\sigma^2_{\epsilon}$, and $\sigma^2_{\textbf{e}}$ for the H-SEM and H-SAM with 90\% and 50\% missing data, respectively. These estimates were obtained using the FML, OML, and MML methods. The table also reports the total computation times for both
parameter estimation and the calculation of standard errors.

\begin{table}[H]
\captionsetup{justification=centering}
\caption{Parameter estimates (est.) along with their standard errors (see.), and computation time in seconds (ct) for fitting the H-SEM model using FML for the full dataset, and OML, MML-D, and MML-P for a partially observed dataset with $90\%$ missing data}
\label{tab:sup_Lucas_SEM_10}
\centering
\begin{tabular}{ccccccccc}
\hline
\multirow{2}{*}{} & \multicolumn{2}{c}{FML}  & \multicolumn{2}{c}{OML} & \multicolumn{2}{c}{MML-D}  & \multicolumn{2}{c}{MML-P}  \\ \cmidrule(lr){2-9}
                           &              est.               &   se.       & est.         & se.      & est. & se.  & est. & se. \\ \hline
            Intercept   & 5.2578            & 0.0748                          & 3.3498& 0.2640 &3.4952 & 0.2446 &3.4952 & 0.2446  \\
age              & 0.6994                          &  0.0793                  & 1.5135& 0.2553 &  1.1073& 0.2472 &  1.1073& 0.2472\\
$\text{age}^2$             & -1.7558                 &         0.1321                  &-3.4992 &0.4678  &-2.6208 & 0.4332 &-2.6208 & 0.4332 \\
$\text{age}^3$        & 0.6355                       &   0.0659                      &  1.1014 & 0.2548 & 0.7277 &0.2287    & 0.7277 &0.2287     \\
log(lotsize) & 0.1458 & 0.0046 & 0.1806& 0.0125 & 0.1743  & 0.0133 & 0.1743  & 0.0133  \\
rooms & 0.0056 & 0.0029 &0.0264 & 0.0128 &0.0021 &0.0110 &0.0021 &0.0110 \\
log(TLA) & 0.6038 & 0.0103 & 0.8340& 0.0411 &  0.8300& 0.0365  &  0.8300& 0.0365 \\
beds & 0.0164 &  0.0043 & -0.0241 & 0.0187 & -0.0049 &  0.0165  & -0.0049 &  0.0165 \\
syear1994 & 0.0365 & 0.0067 &0.0660 & 0.0299  & 0.0770& 0.0258  & 0.0770& 0.0258 \\
syear1995 & 0.0799 & 0.0066 & 0.0862& 0.0294  & 0.1206& 0.0253   & 0.1206& 0.0253 \\
syear1996 & 0.0962 & 0.0064 & 0.0977&  0.0287 & 0.1384 & 0.0247 & 0.1384 & 0.0247  \\
syear1997 & 0.1413 & 0.0063 & 0.1386&  0.0277 & 0.1744 &  0.0242 & 0.1744 &  0.0242\\
syear1998 & 0.1937 & 0.0065 & 0.1575& 0.0293  &0.2148 &0.0253  &0.2148 &0.0253\\
$\rho$ & 0.9866 & 0.0002 &  0.6866& 0.0012 & 0.9936& 0.0006 & 0.9936& 0.0006\\
$\sigma^2_{\textbf{e}}$ & 0.0004 & 0.0001 & 0.1643 & 0.0030 & 0.0001& 0.0002 & 0.0001& 0.0002\\
$\sigma^2_{\epsilon}$ &  0.0685 & 0.0007 & 0.0001 & 0.0081 & 0.0837 & 0.0035 & 0.0837 & 0.0035\\
\hline 

                   ct &  \multicolumn{2}{c}{ 20.098} & \multicolumn{2}{c}{  1.39} & \multicolumn{2}{c}{2792.409} & \multicolumn{2}{c}{38.65
}  \\
 \hline       
\end{tabular}
\end{table}

%Parameter estimates (est) with their standard errors (se), computation time in seconds (ct) for fitting H-SAM and H-SEM using FML, OML, MML-D, and MML-P for the dataset with $90\%$ missing data.

\begin{table}[H]
\captionsetup{justification=centering}
\caption{
Parameter estimates (est.) along with their standard errors (se.), and computation time in seconds (ct) for fitting the H-SAM model using FML for the full dataset, and OML, MML-D, and MML-P for a partially observed dataset with $90\%$ missing data}
\label{tab:sup_Lucas_SAM_10}
\centering
\begin{tabular}{ccccccccc}
\hline
\multirow{2}{*}{} & \multicolumn{2}{c}{FML}  & \multicolumn{2}{c}{OML} & \multicolumn{2}{c}{MML-D}& \multicolumn{2}{c}{MML-P}\\ \cmidrule(lr){2-9} 
                           &              est.               &   se.       & est.         & se.      & est. & se.   & est. & se.\\
                           \hline
           Intercept   & -0.1124             & 0.0507                          & 3.2539 &0.2683 & 0.1863 & 0.1511 & 0.1863 & 0.1511   \\
age              & 0.9565                         &  0.0429                  & 1.6800&0.2551 & 0.8215 &0.1219  & 0.8215 &0.1219\\
$\text{age}^2$             & -1.5790                &          0.0797                   & -3.7745 &0.4711 & -1.2960 & 0.2327 & -1.2960 & 0.2327 \\
$\text{age}^3$        & 0.3697                       &   0.0440                   &  1.1916 & 0.2583& 0.0989 &0.1264 & 0.0989 &0.1264 \\
log(lotsize) & 0.0413 & 0.0022 &0.1701 &0.0124 & 0.0414 &0.0064  & 0.0414 &0.0064\\
rooms & -0.0052 & 0.0026 & 0.0296& 0.0134& -0.0042 &0.0082 & -0.0042 &0.0082 \\
log(TLA) & 0.4454 & 0.0083 &0.8581 &  0.0422 & 0.5234 &0.0324 & 0.5234 &0.0324\\
beds &  0.0129 & 0.0039 & -0.0312&0.0194 & -0.0143 & 0.0125 & -0.0143 & 0.0125\\
syear1994 & 0.0357 & 0.0066 &0.0620 &0.0311 &  0.0631 &0.0213  &  0.0631 &0.0213  \\
syear1995 & 0.0710 & 0.0064 &0.0900 & 0.0305& 0.0816& 0.0207 & 0.0816& 0.0207\\
syear1996 &  0.0864 & 0.0063 &0.1005 & 0.0297&0.1010 &0.0201 &0.1010 &0.0201  \\
syear1997 &  0.1191 & 0.0062 & 0.1341& 0.0290 &0.1288 & 0.0200 &0.1288 & 0.0200 \\
syear1998 & 0.1675 & 0.0064 & 0.1629 &0.0304 &0.1518 & 0.0212  &0.1518 & 0.0212 \\
$\rho$ & 0.6727 & 0.0001 & 0.0027&0.0311 &  0.6046 & 0.0201  &  0.6046 & 0.0201 \\
$\sigma^2_{\textbf{e}}$ & 0.0399 & 0.0008 &0.0882 &0.0114 &0.0682 & 0.0070  &0.0682 & 0.0070 \\
$\sigma^2_{\epsilon}$ & 0.042 & 0.0009 &  0.0882 &0.0234 & 0.0203& 0.0080 & 0.0203& 0.0080
\\ \hline 
             ct & \multicolumn{2}{c}{12.245} & \multicolumn{2}{c}{1.08} &\multicolumn{2}{c}{1196.3}  & \multicolumn{2}{c}{22.53}  \\ \hline   
\end{tabular}
\end{table}

\begin{table}[H]
\captionsetup{justification=centering}
\caption{Parameter estimates (est.) along with their standard errors (se.), and computation time in seconds (ct) for fitting the H-SEM model using FML for the full dataset, and OML, MML-D, and MML-P for a partially observed dataset with $50\%$ missing data}
\label{tab:sup_Lucas_SEM_50}
\centering
\begin{tabular}{ccccccccc}
\hline
\multirow{2}{*}{} & \multicolumn{2}{c}{FML}  & \multicolumn{2}{c}{OML} & \multicolumn{2}{c}{MML-D}& \multicolumn{2}{c}{MML-P}  \\ \cmidrule(lr){2-9}
                           &              est.               &   se.       & est.         & se.      & est. & se. & est. & se.\\ \hline
                        Intercept   & 5.2578            & 0.0748                          &  3.4785 & 0.1166 &4.9306 &0.1073 &4.9306 &0.1073   \\
age              & 0.6994                          &  0.0793                   & 1.9323 & 0.1097  & 0.8120 & 0.1107 & 0.8120 & 0.1107\\
$\text{age}^2$             & -1.7558                 &         0.1321                & -4.1386 & 0.1913  & -2.0756 & 0.1855 & -2.0756 & 0.1855 \\
$\text{age}^3$        & 0.6355                       &   0.0659                       &  1.4958 & 0.1000  & 0.7802 & 0.0928 & 0.7802 & 0.0928  \\
log(lotsize) & 0.1458 & 0.0046  & 0.2034 & 0.0060  & 0.1627 & 0.0064  & 0.1627 & 0.0064 \\
rooms & 0.0056 & 0.0029  & 0.0130 & 0.0051  & 0.0119 & 0.0042 & 0.0119 & 0.0042\\
log(TLA) & 0.6038 & 0.0103 &   0.7737 & 0.0175  & 0.6277 & 0.0150  & 0.6277 & 0.0150  \\
beds & 0.0164 &  0.0043  & 0.0013 & 0.0077  &  0.0116 & 0.0064 &  0.0116 & 0.0064  \\
syear1994 & 0.0365 & 0.0067  & 0.0448 & 0.0123  &  0.0387 & 0.0101 &  0.0387 & 0.0101\\
syear1995 & 0.0799 & 0.0066 & 0.0982 & 0.0121  &  0.0800 & 0.0099 &  0.0800 & 0.0099 \\
syear1996 & 0.0962 & 0.0064 &  0.1002 & 0.0116  &  0.0889 & 0.0095 &  0.0889 & 0.0095  \\
syear1997 & 0.1413 & 0.0063 & 0.1449 &0.0115  & 0.1362 & 0.0094  & 0.1362 & 0.0094\\
syear1998 & 0.1937 & 0.0065 &  0.1971& 0.0119  &  0.1864 & 0.0097 &  0.1864 & 0.0097 \\
$\rho$ & 0.9866 & 0.0002 & 0.6759 & 0.0168  & 0.9907 & 0.0006 & 0.9907 & 0.0006\\
$\sigma^2_{\textbf{e}}$ & 0.0004 & 0.0001 &  0.0242 & 0.0044  & 0.0002 & 0.0001  & 0.0002 & 0.0001 \\
$\sigma^2_{\epsilon}$ &  0.0685 & 0.0007 & 0.1098 & 0.0055  & 0.0725 & 0.0011  & 0.0725 & 0.0011 \\ \hline    
                   ct & \multicolumn{2}{c}{20.098} & \multicolumn{2}{c}{4.3} & \multicolumn{2}{c}{2248.094} & \multicolumn{2}{c}{49.14}\\
\hline  
\end{tabular}
\end{table}

\begin{table}[H]
\captionsetup{justification=centering}
\caption{Parameter estimates (est.) along with their standard errors (se.), and computation time in seconds (ct) for fitting the H-SAM model using FML for the full dataset, and OML, MML-D, and MML-P for a partially observed dataset with $50\%$ missing data}
\label{tab:sup_Lucas_SAM_50}
\centering
\begin{tabular}{ccccccccc}
\hline
\multirow{2}{*}{} & \multicolumn{2}{c}{FML}  & \multicolumn{2}{c}{OML} & \multicolumn{2}{c}{MML-D} & \multicolumn{2}{c}{MML-P}   \\ \cmidrule(lr){2-9}
                           &              est.               &   se.       & est.         & se.      & est. & se. & est. & se.\\ \hline
  Intercept   & -0.1124             & 0.0507                           & 2.8816 & 0.1229  & -0.0169 & 0.0588 & -0.0169 & 0.0588  \\
age              & 0.9565                         &  0.0429                    &  2.1285 & 0.1073  & 0.7853 & 0.0467 & 0.7853 & 0.0467  \\
$\text{age}^2$             & -1.5790                &          0.0797                     &  -4.4020 & 0.1931  &  -1.2922 & 0.0878 &  -1.2922 & 0.0878\\
$\text{age}^3$        & 0.3697                       &   0.0440                     & 1.4320 & 0.1038  & 0.2805 & 0.0490 & 0.2805 & 0.0490 \\
log(lotsize) & 0.0413 & 0.0022   &  0.1766 & 0.0058 &  0.0340 & 0.0024  &  0.0340 & 0.0024 \\
rooms & -0.0052 & 0.0026   &  0.0148 & 0.0059  & -0.0012 & 0.0033 & -0.0012 & 0.0033  \\
log(TLA) & 0.4454 & 0.0083   & 0.8971 & 0.0193  &  0.3686 &0.0099 &  0.3686 &0.0099   \\
beds &  0.0129 & 0.0039   & -0.0216 & 0.0088 & 0.0031 & 0.0049  & 0.0031 & 0.0049 \\
syear1994 & 0.0357 & 0.0066   & 0.0477 & 0.0145  & 0.0305 & 0.0085  & 0.0305 & 0.0085  \\
syear1995 & 0.0710 & 0.0064   &  0.0884 &0.0141  & 0.0576 & 0.0082 & 0.0576 & 0.0082\\
syear1996 &  0.0864 & 0.0063   & 0.1083 & 0.0136  & 0.0663 & 0.0080  & 0.0663 & 0.0080\\
syear1997 &  0.1191 & 0.0062   &  0.1415 & 0.0135  & 0.0903 & 0.0078  & 0.0903 & 0.0078 \\
syear1998 & 0.1675 & 0.0064   &  0.2014 & 0.0139  & 0.1327 & 0.0082 & 0.1327 & 0.0082\\
$\rho$ & 0.6727 & 0.0001   & 0.0015 & 0.0300 & 0.7235 &    0.0051 & 0.7235 &    0.0051\\
$\sigma^2_{\textbf{e}}$ & 0.0399 & 0.0008   & 0.0901  & 0.0062  &  0.0337 & 0.0013 &  0.0337 & 0.0013   \\
$\sigma^2_{\epsilon}$ & 0.042 & 0.0009   & 0.0901 & 0.0016 &  0.0468 & 0.0012&  0.0468 & 0.0012
\\ \hline  
                  computing time (s) & \multicolumn{2}{c}{ 12.245} & \multicolumn{2}{c}{3.8} & \multicolumn{2}{c}{ 1344.56}  & \multicolumn{2}{c}{26.75 }\\
                  \hline   
\end{tabular}
\end{table}

In Tables~\ref{tab:sup_Lucas_SEM_10} to ~\ref{tab:sup_Lucas_SAM_10}, both MML-D and MML-P produce the same estimates, which are consistently closer to the FML estimates, regarded as the true parameter values, when compared to the OML estimates. This pattern is observed across both models and for both levels of missing data. It is important to note that MML-D necessitates more computation time than MML-P for estimating parameters and calculating standard errors in both models across all percentages of missing data.

\section{Additional proofs for the marginal ML method}
\label{sec:sup_MarginalMLproofs}
%\subsection{Maximum Likelihood estimators of the parameters \texorpdfstring{$\boldsymbol{\beta}$}{} and \texorpdfstring{${\omega}$}{}}

\subsection{Closed form solutions for $\boldsymbol{\beta}$ and $\omega$}

This section derives the analytical forms of ML estimators for $\boldsymbol{\beta}$ and $\omega$ for the proposed marginal ML method presented in Section~\ref{sec:direct.marginal.likelihood} of the main paper.
By differentiating the marginal log-likelihood for $\textbf{y}_o$ in Equation~\eqref{eq:maginallik_me} in the main paper with respect to $\boldsymbol{\beta}$ we get
\begin{equation}
\label{eq:dwrbeta_marginalML}
\begin{split}
    \diffp {\textrm{log} f(\textbf{y}_o;\omega,\boldsymbol{\theta},\rho,\boldsymbol{\beta})}{\boldsymbol{\beta}} &=-\frac{1}{2\omega}\diffp{(\textbf{r}_{o}^\top\textbf{V}_{oo}^{-1}\textbf{r}_{o})}{\boldsymbol{\beta}},
\end{split}
\end{equation}
\noindent by substituting $\textbf{r}_o=\textbf{y}_o-\Tilde{\textbf{X}}_o\boldsymbol{\beta}$,
\begin{equation}
\label{eq:dwrbeta}
\begin{split}
    \diffp {\textrm{log} f(\textbf{y}_o;\omega,\boldsymbol{\theta},\rho,\boldsymbol{\beta})}{\boldsymbol{\beta}} &=-\frac{1}{2\omega}\diffp{(\textbf{y}_o-\tilde{\textbf{X}}_o\boldsymbol{\beta})^\top\textbf{V}_{oo}^{-1}(\textbf{y}_o-\tilde{\textbf{X}}_o\boldsymbol{\beta})}{\boldsymbol{\beta}}\\
    &=-\frac{2}{2\omega}(-\tilde{\textbf{X}}_o)^\top\textbf{V}_{oo}^{-1}(\textbf{y}_o-\tilde{\textbf{X}}_o\boldsymbol{\beta})\\
    &=\frac{1}{\omega}(\tilde{\textbf{X}}_o^\top\textbf{V}_{oo}^{-1}\textbf{y}_o-\tilde{\textbf{X}}_o^\top\textbf{V}_{oo}^{-1}\tilde{\textbf{X}}_o\boldsymbol{\beta}).\\
\end{split}
\end{equation}
Then, by setting Equation~\eqref{eq:dwrbeta} to zero, we obtain the ML estimator for $\boldsymbol{\beta}$ 
\begin{equation}
    \label{eq:betahat}
    \begin{split}
    0&=\frac{1}{\omega}(\tilde{\textbf{X}}_o^\top\textbf{V}_{oo}^{-1}\textbf{y}_o-\tilde{\textbf{X}}_o^\top\textbf{V}_{oo}^{-1}\tilde{\textbf{X}}_o\boldsymbol{\beta})\\
     \tilde{\textbf{X}}_o^\top\textbf{V}_{oo}^{-1}\tilde{\textbf{X}}_o\boldsymbol{\beta}&=\tilde{\textbf{X}}_o^\top\textbf{V}_{oo}^{-1}\textbf{y}_o\\
    \hat{\boldsymbol{\beta}}(\rho,\theta) &=\left(\tilde{\textbf{X}}_o^\top\textbf{V}_{oo}^{-1}\tilde{\textbf{X}}_o\right)^{-1}\tilde{\textbf{X}}_o^\top\textbf{V}_{oo}^{-1}\textbf{y}_o.
    \end{split}
\end{equation}
Similarly, the ML estimator for $\omega$ is derived by differentiating~\eqref{eq:maginallik_me} with respect to $\omega$ as
\begin{equation}
\label{eq:dwromega}
\begin{split}
    \diffp {\textrm{log} f(\textbf{y}_o;\omega,\boldsymbol{\theta},\rho,\boldsymbol{\beta})}{\omega} &=-\frac{n_o}{2\omega}+\frac{1}{2\omega^{2}}\textbf{r}_{o}^\top\textbf{V}_{oo}^{-1}\textbf{r}_{o}.
\end{split}
\end{equation}
By setting Equation~\eqref{eq:dwromega} to zero, the ML estimator for $\hat{\omega}$ is
\begin{equation}
    \label{eq:omegahat}
    \hat{\omega}(\rho,\theta)=\frac{\textbf{r}_{o}^\top\textbf{V}_{oo}^{-1}\textbf{r}_{o}}{n_o}.
\end{equation}

\subsection{Calculate $\text{log}|\textbf{V}_{oo}|$ in the parameterisation approach}
\label{sec:sup_log_det}

From Equation~\eqref{eq:V_oo1} in the main paper, we know that sub-matrix $\textbf{V}_{oo}$ can be written as
\begin{equation}
    \label{eq:V_oo2_sup}
    \textbf{V}_{oo}=\textbf{I}_o+\theta \textbf{B}_o  (\textbf{A}^\top\textbf{A})^{-1} \textbf{B}_o^{\top}.
\end{equation}

Using the matrix determinant lemma~\citep{DING20071223}, the determinant of $\textbf{V}_{oo}$ in Equation~\ref{eq:V_oo2_sup} can be obtained as
\begin{equation}
    \label{eq:V_oodet_sup}
    |\textbf{V}_{oo}|=  |\textbf{A}^\top\textbf{A}+\theta\textbf{B}_o^{\top}\textbf{B}_o||(\textbf{A}^\top\textbf{A})^{-1}||\textbf{I}_o|,
\end{equation}

\noindent and by taking the logarithm of both sides, we get

\begin{equation}
    \label{eq:log_V_oodet_sup}
\begin{split}
        \text{log}|\textbf{V}_{oo}|&=  \text{log}|\textbf{A}^\top\textbf{A}+\theta\textbf{B}_o^{\top}\textbf{B}_o|-\text{log}|\textbf{A}^\top\textbf{A}|+\text{log}|\textbf{I}_o|\\
        &=\text{log}|\textbf{A}^\top\textbf{A}+\theta\textbf{B}_o^{\top}\textbf{B}_o|-\text{log}|\textbf{A}^\top\textbf{A}|.\\
\end{split}
\end{equation}

\section{Computing Information matrix}
\label{sec:info}

To obtain variances for the parameters of interest $\boldsymbol{\beta}, \rho, \sigma_{\epsilon}^2$
and $\sigma^2_{\textbf{e}}$, the standard procedure requires the calculation of either the expected or the observed
information matrix. They are obtained by first calculating the second derivatives of the negative of marginal log-likelihood. The negative of marginal log-likelihood is 
\begin{equation} 
\label{eq:negativemarginal}
    -L_o=-\text{log}f(\textbf{y}_o;\sigma^2_{\epsilon},\sigma^2_\textbf{e},\rho,\boldsymbol{\beta})=\frac{n_o}{2}\textrm{log}(2\pi)+\frac{n_o}{2}\textrm{log}(\sigma^2_{\epsilon})-\frac{1}{2}\textrm{log}|\textbf{V}^{-1}_{oo}|+\frac{1}{2\sigma^2_{\epsilon}}\textbf{r}_o^\top\textbf{V}^{-1}_{oo}\textbf{r}_o,
\end{equation}

\noindent and for the convenience of notation, we denote the negative marginal log-likelihood $-L_{o}=\Bar{L}_{o}$.

\vspace{0.2cm}

First, we discuss the H-SEM. The second derivative of $\Bar{L}_{o}$ with respect to $\boldsymbol{\beta}$
\begin{equation}
    \label{eq:gradbeta_SEM}
    \begin{split}
        \frac{\partial \Bar{L}_{o}}{\partial \boldsymbol{\beta}}&=\frac{1}{2\sigma^2_{\epsilon}}2(-\tilde{\textbf{X}}_o^\top)\textbf{V}_{oo}^{-1}(\textbf{y}_o-\tilde{\textbf{X}}_o\boldsymbol{\beta})\\
        &=-\frac{1}{\sigma_{\epsilon}^2}\tilde{\textbf{X}}_o^\top\textbf{V}_{oo}^{-1}\textbf{r}_o,
    \end{split}
\end{equation}
and since $E(\textbf{r})=E(\textbf{y}_o-{\textbf{X}}_{o}\boldsymbol{\beta})=0$, it is straightforward to show that
\begin{alignat*}{2}
    E\left(\frac{\partial^2 \Bar{L}_{o} }{\partial \boldsymbol{\beta}\partial \rho}\right) = 0, &\qquad
    E\left(\frac{\partial^2 \Bar{L}_{o} }{\partial \boldsymbol{\beta}\partial \sigma^2_{\epsilon}}\right) = 0, &\qquad
    E\left(\frac{\partial^2 \Bar{L}_{o} }{\partial \boldsymbol{\beta}\partial \sigma^2_{\textbf{e}}}\right) = 0,
\end{alignat*}
meaning that $\hat{\boldsymbol{\beta}}$ is asymptotically independent from $\boldsymbol{\zeta}=(\rho,\sigma^2_{\epsilon},\sigma^2_{\textbf{e}})^\top$. Further,
\begin{equation}
    \label{eq:hess.betas}
        E\left(\frac{\partial \Bar{L}_{o}}{\partial \boldsymbol{\beta} \partial\boldsymbol{\beta}^\top}\right)=\frac{1}{\sigma_{\epsilon}^2}\tilde{\textbf{X}}_o^\top\textbf{V}_{oo}^{-1}\tilde{\textbf{X}}_o.
\end{equation}
These results leads to $\sigma_{\epsilon}^2\text{Cov}(\boldsymbol{\hat{\beta}})=\left(\tilde{\textbf{X}}_o^\top\textbf{V}_{oo}^{-1}\tilde{\textbf{X}}_o\right)^{-1}$, and $\text{Cov}(\boldsymbol{\hat{\zeta}})=\left[E\left(\frac{\partial^2\Bar{L}_{o}}{\partial \boldsymbol{\zeta} \partial \boldsymbol{\zeta}^\top}\right)\right]^{-1}$. To calculate $\text{Cov}(\boldsymbol{\hat{\zeta}})$, we recommend utilising the observed information matrix through numerical differentiation. This approach helps prevent the need for frequent and expensive inversions.
Now, we consider the H-SAM, which is more complicated because $\textbf{r}_{o}$ depends on $\rho$. 
The first derivative of $\Bar{L}_{o}$ with respect to $\boldsymbol{\beta}$ is
\begin{equation}
    \label{eq:gradbeta_SAM}
    \begin{split}
        \frac{\partial \Bar{L}_{o}}{\partial \boldsymbol{\beta}}&=\frac{1}{2\omega}2(-{(\textbf{A}^{-1}{\tilde{\textbf{X}}})_{o}}^\top)\textbf{V}_{oo}^{-1}(\textbf{y}_o-(\textbf{A}^{-1}{\tilde{\textbf{X}}})_{o}\boldsymbol{\beta})\\
        &=-\frac{1}{\sigma^2_{\epsilon}}({\textbf{A}^{-1}{\tilde{\textbf{X}}})_{o}}^\top\textbf{V}_{oo}^{-1}\textbf{r}_o.
    \end{split}
\end{equation}
Then, it is straightforward to show that
\begin{equation}
    \label{eq:hess.betas_SAM}
        E\left(\frac{\partial \Bar{L}_{o}}{\partial \boldsymbol{\beta} \partial\boldsymbol{\beta}^\top}\right)=\frac{1}{\sigma_{\epsilon}^2}({\textbf{A}^{-1}{\tilde{\textbf{X}}})_{o}}^\top\textbf{V}_{oo}^{-1}({\textbf{A}^{-1}{\tilde{\textbf{X}}})_{o}}.
\end{equation}
In addition, while $E\left(\frac{\partial^2 \Bar{L}_{o}}{\partial \boldsymbol{\beta}\partial \rho}\right) \neq 0$ , it can be shown that $E\left(\frac{\partial^2 \Bar{L}_{o} }{\partial \boldsymbol{\beta}\partial \sigma^2_{\epsilon}}\right) = E\left(\frac{\partial^2 \Bar{L}_{o} }{\partial \boldsymbol{\beta}\partial \sigma^2_{\textbf{e}}}\right) = 0$, with
\begin{equation}
        E\left(\frac{\partial^2 \Bar{L}_{o}}{\partial \boldsymbol{\beta}\partial \rho}\right)=\frac{1}{\sigma^2_{\epsilon}}\left((\textbf{A}^{-1}\tilde{\textbf{X}})_{o}^\top \textbf{V}_{oo}^{-1}\frac{\partial(\textbf{A}^{-1}\tilde{\textbf{X}})_{o}}{\partial \rho}\boldsymbol{\beta} \right),
\end{equation}

\noindent where $\frac{\partial(\textbf{A}^{-1}\tilde{\textbf{X}})_{o}}{\partial \rho}=(\textbf{A}^{-1}\textbf{W}\textbf{A}^{-1}\tilde{\textbf{X}})_{o}$.
We recommend employing numerical differentiation to compute the corresponding second derivative for the term $E\left(\frac{\partial^2\Bar{L}_{o}}{\partial \boldsymbol{\zeta} \partial \boldsymbol{\zeta}^\top}\right)$. This allows us to derive the observed information matrix without utilising expensive matrix computations.

\end{document}